\documentclass[a4paper,twocolumn,amsmath,amssymb,nofootinbib,superscriptaddress,accepted=2023-11-28]{quantumarticle}
\pdfoutput=1

\usepackage[pdftex]{graphicx}
\usepackage{dcolumn}
\usepackage{bm}
\usepackage{amsmath}
\usepackage{indentfirst}
\usepackage{float}
\usepackage[colorlinks]{hyperref}
\usepackage[dvipsnames]{xcolor}
\usepackage{xcolor}
\usepackage{subfigure}
\usepackage{algorithm}
\usepackage{algorithmicx}
\usepackage{algpseudocode}
\usepackage{amsmath}
\usepackage{verbatim}
\usepackage{makecell}
\usepackage[normalem]{ulem}

\usepackage[numbers, sort&compress]{natbib}

\newcommand{\Hop}{\hat{H}}
\newcommand{\Pop}{\hat{P}}
\newcommand{\Top}{\hat{T}}

\newcommand{\aop}{\hat{a}}
\newcommand{\adop}{\hat{a}^{\dagger}}
\newcommand{\thetav}{\vec{\theta}}
\newcommand{\SVD}{{\rm SVD}}
\newcommand{\hc}{{\rm H.c.}}
\newcommand{\cpx}{\mathcal{C}}
\newcommand{\cpxpsr}{\mathcal{C}^{{\rm PSR}}}
\newcommand{\cpxbp}{\mathcal{C}^{{\rm BP}}}
\newcommand{\rerror}{\mathcal{R}}
\newcommand{\real}{{\rm Re}}
\newcommand{\grad}{\bold{g}}
\newcommand{\gradbp}{\bold{g}^{{\rm BP}}}
\newcommand{\gradpsr}{\bold{g}^{{\rm PSR}}}
\newcommand{\im}{{\rm i}}

\newcommand{\gcc}[1]{{{#1}}}

\begin{document}

%\preprint{APS/123-QED}

%\bibliographystyle{model1a-num-names}
%\biboptions{square,numbers,comma,sort&compress}

% \title{VQE-MPS: a highly parallelized variational quantum computational chemistry simulator using matrix product states}
\title{Differentiable matrix product states for simulating variational quantum computational chemistry}
\author{Chu Guo}
\thanks{These authors contribute equally}
% \email{guochu604b@gmail.com}
\affiliation{Henan Key Laboratory of Quantum Information and Cryptography, Zhengzhou,
Henan 450000, China}
% \affiliation{\hnu}

%\author{Yingxiang Gao }
%\affiliation{State Key Laboratory of Computer Architecture, Institute of Computing Technology, Chinese Academy of Sciences, Beijing}
\author{Yi Fan}
\thanks{These authors contribute equally}
\affiliation{Hefei National Laboratory for Physical Sciences at the Microscale, University of Science and Technology of China, Hefei, Anhui 230026, China}

\author{Zhiqian Xu}
% \thanks{These authors contribute equally}
\affiliation{Institute of Computing Technology, Chinese Academy of Sciences, Beijing}

\author{Honghui Shang}
%\email{shanghui.ustc@gmail.com}
\email{shanghui.ustc@gmail.com}
\affiliation{Key Laboratory of Precision and Intelligent Chemistry,University of Science and Technology of China, Hefei,Anhui 230026, China}

%\author{Zhenyu Li}
%\affiliation{Hefei National Laboratory for Physical Sciences at the Microscale, University of Science and Technology of China, Hefei, Anhui 230026, China}

% \author{Chu Guo}
% \email{guochu604b@gmail.com}
% \affiliation{Henan Key Laboratory of Quantum Information and Cryptography, Zhengzhou,
% Henan 450000, China}
% \affiliation{\hnu}

%\date{\today}

\begin{abstract}

Quantum Computing is believed to be the ultimate solution for quantum chemistry problems. Before the advent of large-scale, fully fault-tolerant quantum computers, the variational quantum eigensolver~(VQE) is a promising heuristic quantum algorithm to solve real world quantum chemistry problems on near-term noisy quantum computers. Here we propose a highly parallelizable classical simulator for VQE based on the matrix product state representation of quantum state, which significantly extend the simulation range of the existing simulators. Our simulator seamlessly integrates the quantum circuit evolution into the classical auto-differentiation framework, thus the gradients could be computed efficiently similar to the classical deep neural network, with a scaling that is independent of the number of variational parameters.
As applications, we use our simulator to study commonly used small molecules such as HF, HCl, LiH and H$_2$O, as well as larger molecules CO$_2$, \gcc{BeH$_2$ and H$_4$ with up to $40$ qubits.} 
% We demonstrate our simulator on molecules whose variational quantum circuits contain up to $XXX$ qubits and $XXX$ two-qubit quantum gate operations, as well as its large scale parallelizability with increasing computational resources. 
The favorable scaling of our simulator against the number of qubits and the number of parameters could make it an ideal testing ground for near-term quantum algorithms and a perfect benchmarking baseline for oncoming large scale VQE experiments on noisy quantum computers. 
\end{abstract}

\keywords{Quantum computing, variational quantum eigensolver, matrix product states, linear scaling, parallel scalability}

\maketitle

%\footnotetext[1]{Both authors contributed equally to this work. }

\section{Introduction}
Computational quantum chemistry is an elementary tool for the material and biochemistry science. However exactly solving the electronic Schr\"odinger equation, namely the full configuration interaction (FCI) method, has a complexity that scales exponentially with the system sizes. As a result the FCI method is strictly limited within $20$ orbitals on classical computers. In the meantime, the fast development of quantum computing technologies already allows to manipulate more than $50$ qubits with moderate circuit depth~\cite{AruteMartinis2019,WuPan2021,ZhuPan2022}. In combination with the variational quantum eigensolver (VQE) which does not require fault-tolerant qubits in principle~\cite{AVQE,DVQE,VQEbenchmark,CaoGuzik2019}, it is possible to solve real quantum chemistry problems on near-term noisy quantum computers~\cite{Google2022a,Google2022b}.

As a heuristic quantum algorithm, the effectiveness of VQE could only be tested in real applications, similar to the variational tensor network states algorithm used to solve quantum many-body problems~\cite{Schollwock2011,Orus2014}. Due to the noisy nature of the current and near-term quantum computers, it is crucial to have a perfect reference to benchmark the precision of the computations performed on those quantum computers. Classical simulators serve for this purpose. Moreover, efficient classical simulators allow researchers to explore the advantages or limitations of a given variational quantum circuit ansatz in absence of an actual quantum computer. Currently, the most commonly used classical simulator for VQE is the brute force method, which directly stores an $N$-qubit quantum state as a state vector of $2^N$ complex numbers~\cite{LaRose2019overviewcomparison,Quest}. Due to the exponential memory requirement, with this method it is extremely difficult to simulate beyond $50$ qubits even with the most powerful existing supercomputer. In fact, most classical simulations of VQE reported in literatures are limited within $30$ qubits~\cite{bylaska2021quantum,YalSenGun21,ManKhaYam21,XiaKai20,LiLv2022,LiuWanLi20,FanLiuLi21,KotSchTam21,CaoHuZha21,RyaIZmGen21,CaoYung2022}.

% As such it is important to have an efficient classical simulator for VQE as 

% On the other hand, quantum computing method is promising way to solve this problem. On today's noisy intermediate-scale quantum (NISQ) devices, the variational quantum eigensolver ~(VQE) method is the most popular hybrid quantum-classical algorithm in the compuational quantum chemistry \cite{AVQE,DVQE,VQEbenchmark,CaoGuzik2019}. 

% The VQE method is usually simulated with state-vector method\cite{LaRose2019overviewcomparison,Quest}, that one holds the full quantum state in memory as a large vector of size $2^N$ (N is the number of qubits), and the required computing memory scales exponentially with N, so it is difficult to simulate a system with more than 50 orbitals~(50 qubit) even with the current leading-edge supercomputers\cite{Li2021}. 

To overcome the memory bottleneck and increase the simulation efficiency, we propose an matrix product states (MPS) based simulator for VQE. MPS is known to be able to efficiently represent a class of low entanglement quantum states satisfying an area law~\cite{Hastings2007}. During the past twenty years, it has become a standard ansatz to represent low energy states of one-dimensional quantum many-body systems~\cite{Schollwock2011}. The MPS based simulator has also been applied for simulating quantum circuits~\cite{McCaskeyHumble2018}, random quantum circuit sampling~\cite{ZhouXavier2020} as well as VQE~\cite{ShangLi2022}, and it has also been integrated into numerical packages such as qiskit~\cite{qiskit} and quimb~\cite{quimb}. However, in terms of simulating variational quantum circuits the existing MPS simulators lack an efficient way to compute the gradients beyond the universal parameter shift rule (PSR)~\cite{MitaraiFujii2018} (the complexity of PSR grows linearly with the number of variational parameters) or large scale parallelizability. The contribution of this work is two fold: 1) we propose a strategy to integrate our MPS based simulator into the classical auto-differentiation framework, such that the complexity of computing the gradients is independent of the number of variational parameters in the circuit ansatz; 2) we propose and implement a parallelization scheme for our simulator on distributed architectures. \gcc{As applications, we calculate the ground state energies of commonly used smaller molecules including HF, HCl, LiH and H$_2$O, reaching comparable accuracies to existing approaches, we also study larger molecules such as CO$_2$, BeH$_2$ and H$_4$ with up to $40$ qubits, distinguishing our method from the commonly used state-vector approach which is mostly limited within $30$ qubits.} The favorable scaling of our simulator against both the number of qubits and the number of parameters could make it a promising tool to study near-term quantum algorithms.

% In order to overcome this bottleneck, in this work, we use tensor network based algorithm, specifically the matrix product states~(MPS) method~\cite{Schollwock2011} to simulate the VQE circuits, and the computing time scales linearly with in the number of qubits $N$\cite{GuoChu2021,Zhou2020}. The MPS method takes advantage of the fact that in NISQ devices, decoherence limits the amount of entanglement. The MPS method approximates the low-entangled quantum states with simpler structures, that only spans a very tiny fraction of the overall Hilbert space, so the required computational resources have been dramatically reduced. Although the MPS method can  significantly reduce the computational memory and time requirement for the simulation of single VQE circuit, the simulation of a whole VQE calculation is still very heavy tasks because of the huge number of quantum ansatz circuits. Hence we further design a parallel simulation algorithm based on distributed memory over the circuits, which just ``mimic'' the actual quantum computers, and our method can offer a good reference for the quantum devices.  Based on these developments, a highly accurate VQE-MPS quantum simulator has been constructed, which provides an attractive quantum simulation tool for solving the quantum many-body problem of realistic chemistry system.  

The paper is organized as follows. In Sec.~\ref{sec:method}, the framework of variational quantum eigensolver for quantum computational chemistry is introduced. In Sec.~\ref{sec:mps} we introduce the implementation of quantum circuit simulator based on MPS and then in Sec.~\ref{sec:ad}, we propose the back propagation algorithm for the MPS based simulator. In Sec.~\ref{sec:result} we first validate our simulator by benchmarking it against other approaches in terms of both accuracy and runtime efficiency, and then we demonstrate several applications for commonly used molecules. We conclude in Sec.~\ref{sec:conclusions}.

 % our approach and implementation is validated by comparing the total energy of various molecular systems to the full CI values, furthermore, we discuss the convergence behavior of our implementation, the parallel performance when a large number of cores is used. Finally, Section~\ref{sec:conclusions} summarizes the main ideas and findings of this work and highlights possible future research directions. 

%The capabilities of the the current generation of quantum computing devices do not allow for solving most of the problems. Improvements in the quantum computing power are crucial for exact understanding of chemical properties, which enables studying non-adiabatic effects of some systems, which is important for the action mechanisms of many enzymes.

%However, solving quantum chemistry problems in NISQ is limited both by hardware and algorithmic implementation(the choice of the good inital configuration, ansatz) 

% \section{Matrix Product States}
% \label{sec:method}
\section{VQE for quantum chemistry} \label{sec:method}

The variational quantum eigensolver encodes the ground state of a quantum Hamiltonian into a variational quantum circuit and then minimize the expectation value of the Hamiltonian against the variational parameters using a hybrid quantum-classical optimization procedure. It is a hardware-friendly alternative compared to the well established quantum phase estimation algorithm, mainly due to its flexibility in choosing different quantum circuit ansatz as well as its possible robustness against certain level of noises in the underlying quantum circuit~\cite{Cerezo2021}. 
The variational quantum circuit ansatz is, in principle, very similar to the tensor network ansatz used in classical variational tensor network algorithms, and its expressive power ultimately determines the power of VQE for a certain Hamiltonian. Compared to classical variational ansatz, a quantum circuit with even a very moderate circuit depth $d$ ($d\propto N$ with $N$ the number of qubits) could be extremely difficult to classically simulate, such as the random quantum circuits~\cite{AruteMartinis2019,WuPan2021,ZhuPan2022}. It has also been shown that a shallow quantum circuit could efficiently generate any MPS~\cite{SchonWolf2005} as well as some specific instances of projected entangled pair states (PEPS)~\cite{WeiCirac2022}. Therefore it is possible that a variational quantum circuit could better represent and thus allow to more efficiently compute the ground states of certain Hamiltonians than commonly used classical ansatz (PEPS is believed to be a powerful ansatz for certain two-dimensional Hamiltonians, but manipulating PEPS on classical computers could be very hard).
% (PEPS is believed to be expressive for certain two-dimensional ground states, but computing expectation values exactly on classical computers is a hard problem).

% The variational quantum
% eigensolver (VQE) is a promising algorithm that can be used
% to obtain eigenstates of many-body systems in chemistry, physics, and material
% science using near-term quantum computers. Different from the quantum phase estimation (QPE), VQE is based
% on the Rayleigh–Ritz variational principle and requires a parametric circuit ansatz to perform optimization. It embeds state preparation circuits and quantum measurements into a classical optimization procedure, therefore leads to a much shallower circuit. In such a hybrid quantum–classical manner, expectation value of the Hamiltonian is evaluated on quantum processors, while circuit parameters are updated on a classical computer using an optimization algorithm. 

In recent years, quantum chemistry calculations for molecular systems using VQE has been demonstrated using numerical simulations on most of the leading quantum computer architectures~\cite{ShangLi2022, peruzzo_variational_2014, omalley_scalable_2016, kandala_hardware-efficient_2017, colless_computation_2018, expri-symmetry-1,expri-vqe-1,hempel_quantum_2018,nam_ground-state_2019}. The electronic Hamiltonian of a chemical system can be written in a second-quantized formulation:
\begin{align}\label{eq:ham-pbc}
\Hop=\sum_{p,q}{h_{q}^{p}\adop_{p}\aop_{q}}+\frac{1}{2}{\sum_{\substack{p,q\\r,s}}}{g_{rs}^{pq}\adop_{p} \adop_{q} \aop_{r}\aop_{s}},
\end{align}
where $p, q, r, s$ label the orbitals, $\adop_p$ and $\aop_p$ denote the fermionic creation and annihilation operators for orbital $p$, $h_{q}^{p}$ and $g_{rs}^{pq}$ are one- and two-electron integrals in molecular orbital basis.
In VQE, the creation and annihilation operators are first mapped to weighted Pauli strings using the Jordan-Wigner or Bravyi-Kitaev transformations, and then the energy can be obtained by summarizing the expectation values of all the Pauli strings:
\begin{align}\label{eq:expect-val}
    E(\thetav)&=\langle \psi(\thetav)|\hat{H}|\psi(\thetav) \rangle \nonumber \\
    &=\sum_{p, q}{\langle\psi(\thetav)|{\tilde{h}_{q}^{p} \Pop_{q}^{p}}|\psi(\thetav) \rangle} + \sum_{\substack{p,q\\r,s}}{\langle \psi(\thetav)| {\tilde{g}_{rs}^{pq}\Pop_{rs}^{pq}}|\psi(\thetav) \rangle},
\end{align}
where the parametric wave function ansatz $|\psi(\thetav) \rangle$ is prepared using a quantum circuit containing optimizable parameters denoted as $\thetav$, $\Pop_{q}^{p}$ and  $\Pop_{rs}^{pq}$ denote the Pauli strings corresponding to the Hamiltonian terms $\adop_{p}\aop_{q}$ and $\adop_{p} \adop_{q} \aop_{r}\aop_{s}$ respectively.

The key ingredient of VQE is the underlying variational quantum circuit ansatz.  
In this work, we will use the physically motivated unitary coupled-cluster (UCC) ansatz~\cite{UCCansatz} (however our approach is completely general for other ansatz), where the wave function takes the form
\begin{align}\label{eq:uccansatz}
    |\Psi\rangle=e^{\Top-\Top^{\dagger}}|\Phi_{0}\rangle
\end{align}
with $|\Phi_{0}\rangle$ a reference state, typically chosen as the Hartree-Fock ground state. $\Top$ is the cluster operator. When truncated to the single and double excitations (UCCSD), we have
\begin{align}
\Top =\sum_{a,i}\theta_{i}^{a}\Top_{i}^{a} +\frac{1}{4}{\sum_{\substack{a,b\\i,j}}} \theta_{i j}^{a b} \Top_{i j}^{a b}
\end{align}
with
\begin{align}
    \Top_{q}^{p} &= \adop_{p} \aop_{q}; \\
    \Top_{r s}^{p q} &= \adop_{p} \adop_{q} \aop_{r} \aop_{s},
\end{align}
where $\{i, j, k, \dots \}$, $\{a, b, c, \dots\}$, and $\{p, q, r, \dots\}$ indicate the occupied, virtual and general orbitals respectively.
Using the Trotter-Suzuki decomposition $e^{\hat{A}+\hat{B}}\approx(e^{\hat{A}/k}e^{\hat{B}/k})^{k}$~\cite{Babbush_Trotter_2015, Grimsley_Trotter_2020}, the operator exponential in the UCC ansatz in Eq.(\ref{eq:uccansatz}) can be converted into a quantum circuit structured as successive V-shaped blocks. Truncating the Trotterization procedure to a finite order $k$ results in a quantum circuit with depth approximately $\mathcal{O}(kN^4)$, where $N$ is the number of qubits. 

\section{MPS based quantum circuit simulator}\label{sec:mps}

\begin{figure}
\centering
  \includegraphics[width=\linewidth]{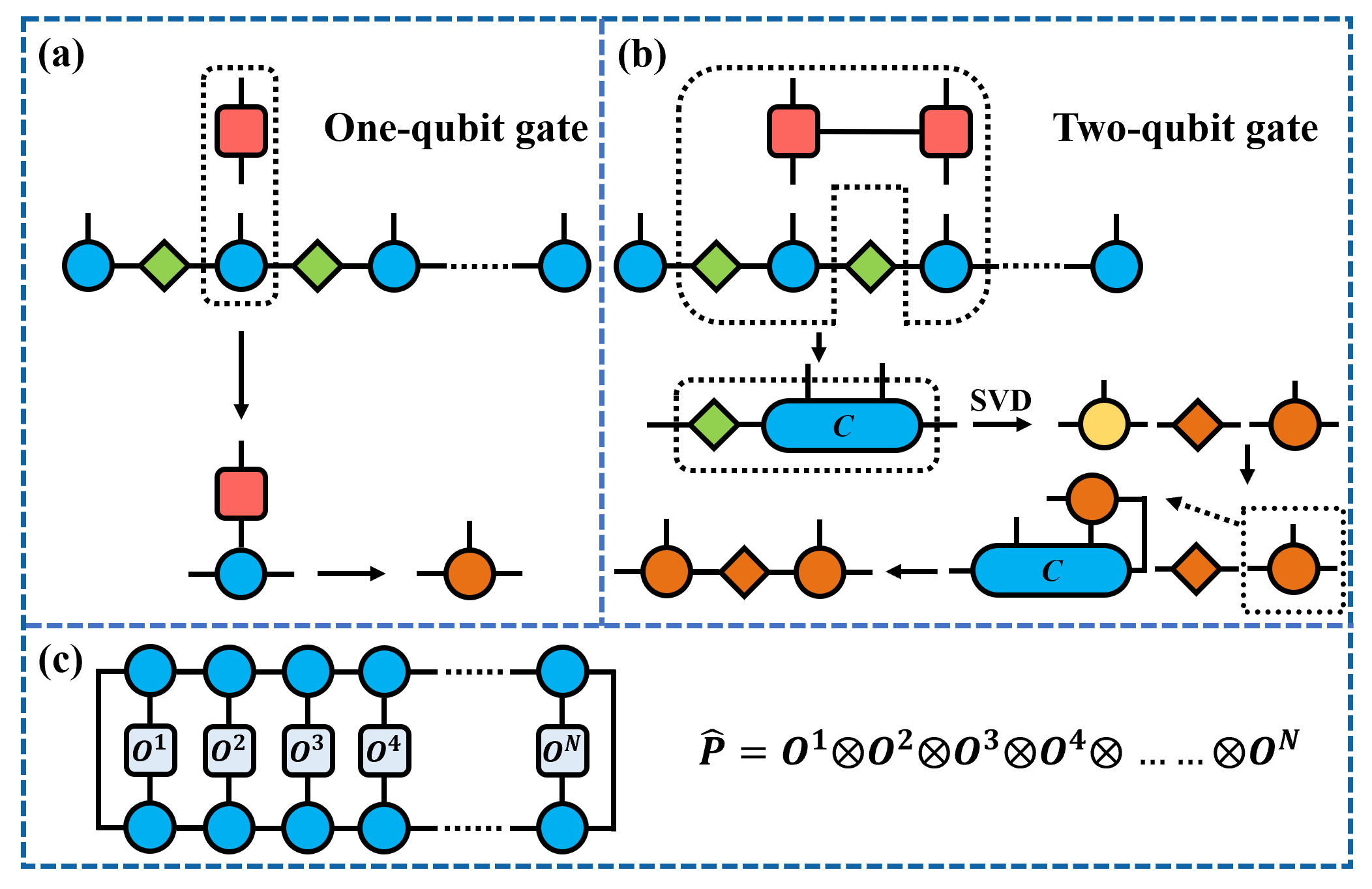}
  \caption{(a) Single-qubit gate operation on an MPS. (b) Nearest-neighbour two-qubit gate operation on an MPS, which preserves the right-canonical form of the MPS. (c) Computing the expectation value of a general Pauli string on an MPS. \gcc{In all these panels, the circles represent the MPS site tensors, the diamonds represent the bipartition singular vectors, and the squares represent the single-qubit and two-qubit operators. In panel (a) and (b), we have used different colors for the same objects to stress that they have been updated during the gate operations.}
  }
  \label{fig:one-and-two-qubits}
\end{figure}

The correlated wave function, generated by the FCI method in quantum chemistry by considering all excitations above the Hartree-Fock reference state, can be written as:

\begin{align}
\left|\Psi\right\rangle  = \sum_{i_1...i_N} c_{i_1 i_2 i_3 \ldots i_N} |i_1 i_2 i_3 \ldots i_N \rangle
\end{align} 
where $|i_1 i_2 i_3 \ldots i_N \rangle$ refers to a specific computation basis, and the coefficients $c_{i_1 i_2 i_3 \ldots i_N}$ constitute a rank-$N$ tensor of $2^N$ complex numbers, which are the parameters to be optimized if the energy is to be minimized in a brute force way (the state-vector approach). This high-rank tensor can also be represented as a matrix product state (MPS), which is the product of a chain of rank-$3$ tensors written as
\begin{align}\label{eq:mps}
c_{i_1 i_2 i_3 \ldots i_N} = \sum_{\alpha_0...\alpha_{N}}  
B^{i_1}_{\alpha_0 \alpha_1} B^{i_2}_{\alpha_1\alpha_2}
B^{i_3}_{\alpha_2\alpha_3} \ldots  B^{i_{N}}_{\alpha_{N-1}\alpha_N},
\end{align}
where $i_n \in \{0,1\}$ refers to ``physical'' index and $\alpha_n$ the ``virtual'' index related to the partition entanglement entropy. The virtual indices $\alpha_0$ and $\alpha_N$ at the boundaries are trivial indices added for notational convenience. The largest size of the virtual indices,
\begin{align}
D = \max_{1\leq n\leq N}\left(\dim (\alpha_n)\right),
\end{align}
is referred to as the \textit{bond dimension} of MPS, which characterizes the memory and computational complexity of MPS based algorithms. Eq.(\ref{eq:mps}) is able to represent any quantum state exactly if $D$ increases exponentially. While a quantum state is said to be efficiently represented as an MPS if $D$ is nearly a constant when $N$ grows.

To simulate VQE a general purpose quantum circuit simulator is required. For tensor network states based simulators three ingredients are usually necessary~\cite{GuoChu2019}: 1) preparing an initial separate quantum state as a tensor network state; 2) performing quantum gate operations onto the initial state and updating the underlying tensor network state; 3) computing expectation values for each Hamiltonian term (namely each term in Eq.(\ref{eq:ham-pbc})). 
% Some of the details of implementing those three functions based on MPS have already been presented in Ref.~\cite{ShangLi2022}. 
In the following we will present the detailed implementation of these functions based on MPS (we note that some functions have already been shown in our previous work~\cite{ShangLi2022}, which will still be included here for self-completeness). These functions have also been used to simulate the Hamiltonian evolution of quantum many-body systems~\cite{Schollwock2011}.

% These ingredients are exactly the same as those required in the time-evolving MPS (t-MPS) algorithms used for time evolution of one-dimensional quantum many-body Hamiltonians~\cite{XXX}. In the following we will briefly summarize a specific t-MPS algorithm used in this work.

The Hartree-Fock reference state is simply a separable state which can be written as 
\begin{align}\label{eq:hf}
\vert \Psi_0\rangle = \vert b_1 b_2 b_3 \ldots b_N\rangle.
\end{align}
Any quantum state in the form of Eq.(\ref{eq:hf}) can be efficiently written as an MPS with $D=1$, by setting each site tensor of the MPS as
\begin{align}
B^{i_n = b_n}_{\alpha_{n-1}=0,\alpha_n=0} = 1, \quad B^{i_n = 1-b_n}_{\alpha_{n-1}=0,\alpha_n=0} = 0.
\end{align}

When simulating the gate operations on MPS, it is important to keep the underlying MPS in a canonical form for accuracy and stability of the algorithm~\cite{Schollwock2011}. We use the right canonical form of the MPS, namely each site tensor $B^{i_n}_{\alpha_{n-1} \alpha_n}$ satisfies the right-canonical condition
\begin{align}\label{eq:rightcanonicalcondition}
\sum_{i_n, \alpha_n} (B^{i_n}_{\alpha_{n-1}' \alpha_n})^{\ast} B^{i_n}_{\alpha_{n-1} \alpha_n} = \delta_{\alpha_{n-1} \alpha_{n-1}'},
\end{align}
and we will perform the gate operations in a way that the right canonical form of MPS is preserved during the evolution (the initial state is naturally right canonical since it is separable), we will also save all the bipartition Schmidt numbers of the MPS similar to the time evolving block decimation (TEBD) algorithm~\cite{Vidal2004}. 
A single-qubit gate operation acting on the $n$-th qubit, denoted as $Q_{i_n}^{i'_n}$ can be simply applied onto the MPS as
\begin{align}
\tilde{B}^{i'_n}_{\alpha_{n-1}\alpha_{n}}  = \sum_{i_n'} Q_{i_n}^{i'_n} B^{i_n}_{\alpha_{n-1}\alpha_{n}}.
\end{align}
The new site tensor $\tilde{B}^{i'_n}_{\alpha_{n-1}\alpha_{n}}$ satisfies Eq.(\ref{eq:rightcanonicalcondition}) as long as $B^{i_n}_{\alpha_{n-1}\alpha_{n}}$ satisfies Eq.(\ref{eq:rightcanonicalcondition}) ($Q_{i_n}^{i'_n}$ is unitary). 
For a nearest-neighbour two-qubit gate acting on two qubits $n$ and $n+1$ (the $n$-th bond), denoted as $Q_{i_ni_{n+1}}^{i'_n i'_{n+1}}$, we use the technique in Ref.~\cite{Hastings2009} to preserve the right canonical form, which is shown as follows. First we contract the two site tensors $B^{i_n}_{\alpha_{n-1}\alpha_n}$ and $B^{i_{n+1}}_{\alpha_n\alpha_{n+1}}$ with $Q_{i_n i_{n+1}}^{i'_n i'_{n+1}}$ to obtain a two-site tensor
\begin{align}
C_{\alpha_{n-1}\alpha_{n+1}}^{i'_n i'_{n+1}}=\sum_{\alpha_n, i_n,i_{n+1}} Q_{i_n i_{n+1}}^{i'_n i'_{n+1}} B^{i_n}_{\alpha_{n-1} \alpha_n} B^{i_{n+1}}_{\alpha_n\alpha_{n+1}}.
\end{align}
Then we contract $C_{\alpha_{n-1}\alpha_{n+1}}^{i'_n i'_{n+1}}$ with the Schmidt numbers stored at the $n-1$-th bond (denoted as $\lambda_{\alpha_{n-1}}$) to obtain a new two-site tensor as
\begin{align}
\tilde{C}_{\alpha_{n-1} \alpha_{n+1}}^{i'_n,i'_{n+1}}= \lambda_{\alpha_{n-1}} C_{\alpha_{n-1} \alpha_{n+1}}^{i'_n i'_{n+1}},
\end{align}
% Usually, to perform a two-qubit gate $Q$ between qubit $n$ and qubit $n+1$, 
% the MPS should be first transformed to ``canonical form'' centered around the qubits of interest, through a series of $QR$ factorizations \cite{Schollwock2011}. Here we keep our MPS in right-canonical form by ensuring that $\sum_{i_n,\sigma_n} (C^{i_n}_{\sigma_{n-1},\sigma_n})^{*}C^{i_n}_{\sigma'_{n-1},\sigma_n}=\delta_{\sigma_{n-1},\sigma'_{n-1}}$. Assuming that the initial MPS state is already in right-canonical form, the two-qubit gate $Q^{i_n,i_{n+1}}_{i'_n,i'_{n+1}}$ can be applied onto the $n$-th bond between tensor $n$ and $n+1$ while maintaining the right-canonical form. As shown in Fig.~\ref{fig:one-and-two-qubits}(b), one obtains the rank-4 tensor $T$ by contracting the two-qubit gate with tensors $C^{i_n}_{\sigma_{n-1},\sigma_n}$ and $C^{i_{n+1}}_{\sigma_n,\sigma_{n+1}}$
% In order to restore the MPS form, a singular value decomposition (SVD) should be performed to $T$ to obtain two rank-3 site tensors, however the right-canonical property of the first rank-3 tensor is no longer satisfied. Instead, the SVD should be performed to
Now we perform singular value decomposition (SVD) onto the tensor $\tilde{C}_{\alpha_{n-1} \alpha_{n+1}}^{i_n' i’_{n+1}}$ and get
% where $T'_{\sigma_{n-1},i_n,i_{n+1},\sigma_{n+1}}$ has center-canonical form (why?), so that we can apply MPS truncation on it in SVD:
\begin{equation}
\SVD(\tilde{C}_{\alpha_{n-1}\alpha_{n+1}}^{i_n' i’_{n+1}})=\sum_{\alpha_n} U^{i’_n}_{\alpha_{n-1} \alpha_n} \tilde{\lambda}_{\alpha_n} V^{i'_{n+1}}_{\alpha_n\alpha_{n+1}},
\end{equation}
during which we will also truncate the small Schmidt numbers below a certain threshold or simply reserve the $D$ largest Schmidt numbers if the number of Schmidt numbers larger than the threshold is still larger than $D$. Now the new site tensors $\tilde{B}^{i'_n}_{\alpha_{n-1}\alpha_n}$ and $\tilde{B}^{i’_{n+1}}_{\alpha_n\alpha_{n+1}}$ can be obtained as
\begin{align}
\tilde{B}^{i'_n}_{\alpha_{n-1}\alpha_n} &= \sum_{i'_{n+1},\alpha_{n+1}} C_{\alpha_{n-1}\alpha_{n+1}}^{i'_n i’_{n+1}} \left(V^{i'_{n+1}}_{\alpha_n\alpha_{n+1}}\right)^{\ast}; \\
\tilde{B}^{i’_{n+1}}_{\alpha_n\alpha_{n+1}} &= V^{i'_{n+1}}_{\alpha_n\alpha_{n+1}}.
\end{align}
$\tilde{B}^{i'_n}_{\alpha_{n-1}\alpha_n}$ and $\tilde{B}^{i’_{n+1}}_{\alpha_n\alpha_{n+1}}$ are indeed the solutions
since $\sum_{\alpha_n} \tilde{B}^{i'_n}_{\alpha_{n-1}\alpha_n} \tilde{B}^{i’_{n+1}}_{\alpha_n\alpha_{n+1}} = C_{\alpha_{n-1}\alpha_{n+1}}^{i'_n i’_{n+1}}$ up to truncation errors. In addition, $\tilde{B}^{i'_{n+1}}_{\alpha_n\alpha_{n+1}}$ satisfies Eq.(\ref{eq:rightcanonicalcondition}) by construction, and $\tilde{B}^{i'_n}_{\alpha_{n-1}\alpha_n}$ also satisfies Eq.(\ref{eq:rightcanonicalcondition}) by verifying that $\tilde{B}^{i'_n}_{\alpha_{n-1}\alpha_n} = U^{i’_n}_{\alpha_{n-1}\alpha_n} \tilde{\lambda}_{\alpha_n}/\lambda_{\alpha_{n-1}} $. The new Schmidt numbers $\tilde{\lambda}_{\alpha_n}$ is used to replace the old $\lambda_{\alpha_n}$ at the $n$-th bond. The application of a single-qubit and a nearest neighbour two-qubit gate are also shown in Fig.~\ref{fig:one-and-two-qubits}(a,b). In practice, one can absorb the single-qubit gate operations into two-qubit gate operations using gate fusion, which could further increase the simulation efficiency. A non-nearest-neighbour two-qubit gate could be realized by decomposing it into a series of nearest-neighbour two-qubits gate using SWAP gate. 

Last, given a single Pauli string denoted as
\begin{align} 
\hat{P} = O^1 \otimes O^2 \otimes \ldots \otimes O^N,
\end{align}
where each $O^i \in \{X, Y, Z, I_2\}$ is a single-qubit observable ($X$, $Y$, $Z$ are the Pauli matrices and $I_2$ is the $2\times 2$ identity matrix), the expectation value of $\hat{P}$ on an MPS can be computed by
\begin{align}\label{eq:paulistringexp}
\langle \Psi\vert \hat{P}\vert \Psi\rangle = \sum_{\substack{i_{N:1}, i'_{N:1}\\ \alpha_{N:0}, \alpha'_{N-1:1}}} & \left(O^{i'_1}_{i_1} B^{i_1}_{\alpha_0 \alpha_1} B^{i'_1}_{\alpha_0 \alpha'_1}\right) \times \cdots \nonumber \\
&  \times \left(O^{i'_N}_{i_N}  B^{i_N}_{\alpha_{N-1} \alpha_N} B^{i'_N}_{\alpha_{N-1} \alpha'_N}\right) ,
\end{align}
where we have used $x_{j:i} = \{x_i, x_{i+1}, \dots, x_j\}$ as an abbreviation for a list of indices. Eq.(\ref{eq:paulistringexp}) amounts to contracting a one-dimensional tensor network with a computational complexity $O(ND^3)$, which is also shown in Fig.~\ref{fig:one-and-two-qubits}(c).

\section{Back propagation algorithm for MPS}\label{sec:ad}

VQE is a minimization problem with loss function in Eq.(\ref{eq:expect-val}). When the loss function contains a large number of tunable parameters, one generally needs to use a gradient-based optimization algorithm to accelerate convergence. The standard way to compute the gradient of Eq.(\ref{eq:expect-val}) exactly (on quantum computers) is to use the parameter shift rule, namely
\begin{align}\label{eq:psr}
\frac{\partial E(\thetav)}{\partial \theta_j} = \frac{1}{2}\left(E(\thetav_j^+) - E(\thetav_j^-)\right),
\end{align}
where $\thetav_j^{\pm}$ means to shift the $j$-th parameter of $\thetav$ by $\pm\frac{\pi}{2}$. Clearly, to compute all the gradients using Eq.(\ref{eq:psr}), one would have to evaluate the loss function in Eq.(\ref{eq:expect-val}) for $2M$ times if $\thetav$ contains a total of $M$ parameters.

Although Eq.(\ref{eq:psr}) seems to be the most efficient way to compute the exact gradients on quantum computers, there are shortcuts to compute gradients of Eq.(\ref{eq:expect-val}) on a classical computer by using \textit{automatic differentiation}. In fact, a central reason for the great success of deep learning is the ability to efficiently compute the gradient of a deep neural network using the reverse-mode automatic differentiation. For a deep neural network, the complexity of evaluating the gradients of a loss function against all the parameters using (the reverse-mode) automatic differentiation (back propagation), is roughly equal to that of evaluating the loss function itself (forward propagation)~\cite{autodiff}.
Briefly speaking, automatic differentiation works by treating the loss function as the composition of many ``elementary'' functions, then by defining an adjoint function for each of the elementary function and memorizing all the intermediate outputs generated by those elementary functions, one could systematically compute the gradients of the loss function by evaluating the adjoint functions in a reverse order~\cite{GuoPoletti2021}. In the context of simulating VQE, if we treat each gate operation as an elementary function and memorize all the intermediate outputs during evaluating Eq.(\ref{eq:expect-val}), then the number of copies of the quantum state would scale linearly with the number of gate operations, which is certainly not impractical. Fortunately, by making use of the fact that the quantum circuit is a product of many small unitary gate operations which are reversible, one could use a memory efficient automatic differentiation scheme where only two copies of the quantum state are required at least. 

In the next we will denote the variational quantum circuit ansatz $\vert \psi(\thetav)\rangle$ as 
\begin{align}\label{eq:circuitevolve}
f(\thetav) = \vert \psi(\thetav)\rangle = U(\thetav) \vert \Psi_0\rangle.
\end{align}
Without loss of generality, we denoted the parametric quantum circuit $U(\thetav)$ as
\begin{align}\label{eq:pqc}
U(\thetav) = U(\theta_M) U(\theta_{M-1}) \dots U(\theta_2) U(\theta_1),  
\end{align}
where each parametric quantum gate $U(\theta_j)$ contains a single parameter $\theta_j$.
Then the derivative of Eq.(\ref{eq:expect-val}) against one of the parameter $\theta_j$ can be computed as~\cite{MitaraiFujii2018}
\begin{align}\label{eq:derivative}
\frac{\partial E(\thetav)}{\partial \theta_j} &= \langle \psi(\thetav) \vert \Hop U_{M:j+1} \dot{U}(\theta_j) U_{j-1:1} \vert \Psi_0\rangle + \hc ,
\end{align}
where we have used $U_{j:i} = U(\theta_j) U(\theta_{j-1})\dots  U(\theta_i)$ and $\dot{U}(\theta_j) = \frac{d U(\theta_j)}{d \theta_j}$. Now we further define
\begin{align}
\vert \Psi_j \rangle &= U_{j:1} \vert \Psi_0\rangle ; \label{eq:Psij} \\
\vert \psi_j \rangle &= U_{M:j+1}^{\dagger} \Hop \vert \psi(\thetav)\rangle \label{eq:psij} ,
\end{align}
where we note that $\vert \psi_j \rangle$ is not a proper quantum state in general since $\Hop$ may not be unitary.
With Eqs.(\ref{eq:Psij}, \ref{eq:psij}), Eq.(\ref{eq:derivative}) can be rewritten as
\begin{align}\label{eq:derivative2}
\frac{\partial E(\thetav)}{\partial \theta_j} = \langle \psi_j\vert \dot{U}(\theta_j) \vert \Psi_{j-1}\rangle + \hc.
\end{align}

To this end we note the the gradients of the loss function in Eq.(\ref{eq:expect-val}) can be directly evaluated on a classical computer using Eq.(\ref{eq:derivative2}). Here we show how to embed the evaluation of Eq.(\ref{eq:derivative2}) into the automatic differentiation framework such that one could easily generalize the loss function in Eq.(\ref{eq:expect-val}), for example, to allow classical controls between quantum gate operations, or to allow classical functions which accept the expectation values of each Pauli strings as inputs (the latter would be very helpful for us as will be shown later). To enable automatic differentiation, we treat the loss function in Eq.(\ref{eq:expect-val}) as a composed function of two elementary functions: 1) the quantum circuit evolution in Eq.(\ref{eq:circuitevolve}) (we treat the whole circuit evolution instead of each gate operation as an elementary function) and 2) computing the expectation value on the final state $\vert \psi(\thetav)\rangle$, namely
\begin{align}\label{eq:expec}
E(\vert \psi(\thetav)\rangle) = \langle \psi(\thetav)\vert \Hop \vert \psi(\thetav)\rangle.
\end{align}
The adjoint function of Eq.(\ref{eq:expec}), denoted as $\tilde{E}$, can be defined using the rules in Ref.~\cite{GuoPoletti2021} \gcc{for a function with complex numbers as input and real output:} 
\begin{align}\label{eq:expecgrad}
\tilde{E}(z) &= 2 \real(z) \frac{\partial E(\vert \psi(\thetav)\rangle)}{\partial \langle \psi(\thetav)\vert} = 2 \real(z) \Hop \vert \psi(\thetav)\rangle \nonumber \\ 
&= 2 \real(z) \vert \psi_M \rangle ,
\end{align}
where $z$ is a scalar passed from the possible outer-level adjoint function. The adjoint function of Eq.(\ref{eq:circuitevolve}) can be defined as
\begin{align}\label{eq:circuitevolvegrad}
\tilde{f} (\langle \phi \vert ) = \langle \phi\vert \frac{\partial U(\thetav)}{\partial \thetav} \vert \Psi_0\rangle ,
\end{align}
where $\langle \phi \vert$ is the gradient passed from Eq.(\ref{eq:expecgrad}), and the gradient against $\vert \Psi_0\rangle$ is neglected since it is a constant \gcc{(The back propagation rules for complex functions, which are used to derive Eqs.(\ref{eq:expecgrad}, \ref{eq:circuitevolvegrad}), can be found in Appendix.~\ref{app:bprules})}. Substituting $\langle \phi \vert = \tilde{E}(z)$ into Eq.(\ref{eq:circuitevolvegrad}), we can see that each element of $\tilde{f} (\langle \phi \vert ) $ is exactly Eq.(\ref{eq:derivative2}) up to a constant scalar $\real(z)$ ($z$ will be chosen to be $1$ if there are no outer-level functions on top of $E(\vert \psi(\thetav)\rangle)$, namely $\frac{\partial E(\thetav)}{\partial \thetav} = \tilde{f}(\tilde{E}(1))$). Eq.(\ref{eq:circuitevolvegrad}) (or equivalently Eq.(\ref{eq:derivative2})) can be evaluated in a memory efficient way with no additional copies of the quantum state on top of $\vert \phi\rangle$ and $\vert \Psi_M\rangle = \vert \psi(\thetav)\rangle$, which is shown in Algorithm.~\ref{alg:ad}.

\begin{algorithm}[H]
\caption{Memory-efficient way to evaluate Eq.(\ref{eq:circuitevolvegrad}), with two input states $\vert \phi\rangle$ and $\vert\Psi_M\rangle$.} \label{alg:ad}
\begin{algorithmic}[1]
\State Initialization: $\vert \Psi \rangle = \vert \Psi_M \rangle$, \gcc{$\vert \phi\rangle=\Hop \vert \Psi_M \rangle$} and $grads = zeros(M)$
\For{$j = M:-1:1$}
    \State $ \vert \Psi \rangle \leftarrow U(\theta_{j})^{\dagger} \vert \Psi\rangle   $
    \State $grads[j] = \real(\langle \phi \vert \dot{U}(\theta_j)\vert \Psi \rangle)  $
    \State $\vert \phi \rangle \leftarrow U(\theta_{j})^{\dagger} \vert \phi \rangle $
\EndFor
\State Return $grads$

\end{algorithmic}
\end{algorithm}

From Algorithm.~\ref{alg:ad}, we can see that during the back propagation one needs to apply the reverse of the quantum circuit $U(\thetav)$ onto the two states $\vert \phi\rangle$ and $\vert\Psi_M\rangle$, therefore ideally the complexity of this algorithm is approximately two times the complexity of the forward propagation if we neglect the complexity of computing the ``expectation value'' $\langle \phi \vert \dot{U}(\theta_j)\vert \Psi \rangle$ (which is reasonable if $M$ is much less than the total number of gate operations). While this is true for the state-vector based simulators (the brute-force approach, for which the memory-efficient back propagation has been implemented in packages such as Yao.jl~\cite{Yao}), for our MPS based simulator it is more complicated due to the ``state'' $\vert \psi_j\rangle$ in Eq.(\ref{eq:psij}). Concretely, taking $j=M$, we get $\vert \psi_M\rangle = \Hop \vert \psi(\thetav)\rangle $, which requires to apply the Hamiltonian $\Hop$ onto the final quantum state $\vert \psi(\thetav)\rangle$. Since $\Hop$ is generally the summation of a large number of Pauli strings, this operation will involve the addition of many states resulting from the application of each Pauli string. While for state-vector based simulator this addition can be implemented with no additional memory overhead (and very little computational overhead if the number of terms in the Hamiltonian, denoted as $N_h$, is much smaller than the number of gate operations, denoted as $N_g$), for MPS based simulator the addition of different MPSs will generally result in an MPS with much larger bond dimension (If one add $k$ MPSs with bond dimension $D$ each, the resulting MPS would have a bond dimension $kD$ in the worst case if the addition is done very accurately~\cite{Schollwock2011}). Therefore in MPS based simulator the gate operations on $\vert \psi_j\rangle$ will either have a larger computational complexity than the gate operations on $\vert\Psi_j\rangle$, or have a lower accuracy.
To balance between the computational complexity and the accuracy, we split $\Hop$ into $m$ different groups labeled by $\Hop^j$ with $1\leq j\leq m$, denoted as $\Hop = \sum_{j=1}^m \Hop_j$, where each group contains $N_h / m$ Pauli strings. Then we compute the gradients for each $\Hop^j$ using the memory efficient back propagation algorithm with a fixed bond dimension $D$, denoted as $\grad^j$, and then the final gradients $\grad$ is the summation of all $\grad^j$, namely 
\begin{align}
\grad = \sum_{j=1}^m \grad^j . 
\end{align}
\gcc{We stress that the above equation always holds for loss function in the form of Eq.(\ref{eq:expec}), independent of how we split $\Hop$, since $E = \langle \psi(\thetav)\vert\Hop\vert \psi(\thetav)\rangle = \sum_{j=1}^m \langle \psi(\thetav)\vert\Hop^j\vert \psi(\thetav)\rangle $ holds in general, even if $\Hop^j$ does not commute with each other. However, taking into consideration that we perform MPS truncation during the back propagation, different ways of splitting $\Hop$, such as the choice of the total number of groups $m$ and the choice of the Pauli strings in each group, may affect the accuracy of the final gradient.}
In the following we will refer to the MPS simulator powered by the memory efficient back propagation algorithm as the \textit{differentiable MPS} simulator.

\begin{figure}
\centering
  \includegraphics[width=\linewidth]{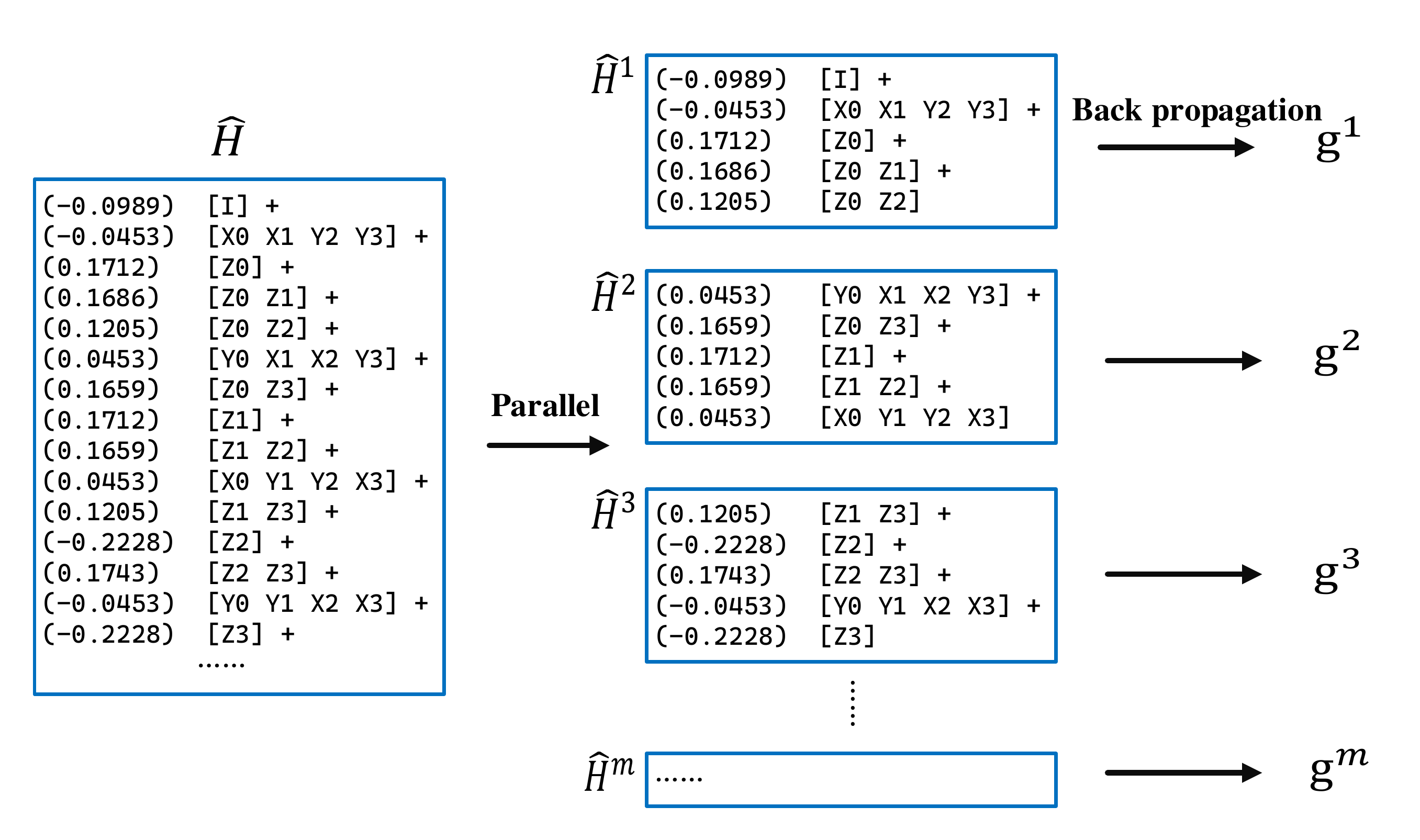}
  \caption{The distribution of Hamiltonian and parallel scheme of gradient evaluation in our work. The Pauli strings (or terms) of global Hamiltonian are equally distributed to MPI processes, with no further data communication required since calculation of expectation value and gradient evaluation completely depend on the local Hamiltonian. The global Hamiltonian containing 15 Pauli strings in the figure is generated from H$_2$ molecule.}
  \label{fig:algorithms}
\end{figure}

Now we analyze the computational complexity of different approaches to compute the gradients based on MPS in detail. We denote the classical complexity of simulating each gate operation as $\cpx_g$ ($\cpx_g = O(D^3)$ for a nearest neighbour two-qubit gate), the complexity of evaluating the expectation value of a single Pauli string as $\cpx_h$ ($\cpx_h = O(ND^3)$ for a general Pauli string). Then the complexity of the forward propagation, denoted as $\cpx$, is $\cpx = N_g \cpx_g + N_h \cpx_h$. The parameter shift rule for computing the gradients has a complexity
\begin{align}\label{eq:cpxpsr}
\cpxpsr = 2M\cpx = 2M(N_g \cpx_g + N_h \cpx_h) .
\end{align}
Assuming that we divide $\Hop$ into $m$ groups, then the complexity of the memory efficient back propagation algorithm for each group is $2 N_g \cpx_g + \frac{N_h}{m} \cpx_h$, and the whole complexity scales as
\begin{align}\label{eq:cpxbp}
\cpxbp = (m+1) N_g \cpx_g + N_h \cpx_h,
\end{align}
where the coefficient of the first term on the right hand side of Eq.(\ref{eq:cpxbp}) is $m+1$ instead of $2m$ since the states $\vert\Psi_j\rangle$ in Eq.(\ref{eq:Psij}) are the same for each group. From Eqs.(\ref{eq:cpxpsr}, \ref{eq:cpxbp}) we can clearly see that our back propagation algorithm is more efficient than PSR if $m \ll M$. 
The overhead occurred in the back propagation algorithm by dividing $\Hop$ into groups could be overcome by parallelization, since the gradients $\grad^j$ for each $\Hop^j$ can be computed almost perfectly in parallel (if one does not share the states $\vert\Psi_j\rangle$ among different processes).
The parallelization strategy to compute the gradients for our differentiable MPS simulator is demonstrated in Fig.~\ref{fig:algorithms}.

To this end, we note that in our approach we have treated the MPS as a ``global'' state similar to the brute force approach. However, when computing the gradient one could also treat MPS ``locally'', which means that one compute the gradient of Eq.(\ref{eq:expect-val}) with respect to each site tensor of the MPS instead of the global state~\cite{LiaoXiang2019}. The possible advantage of this approach is that one could in principle find the exact gradient with respect to those site tensors and then optimize each of them within a fixed bond dimension $D$, while in our approach the gradient would no longer be accurate if the intermediate states during the back propagation could no longer be written as an MPS with bond dimension $D$. However, in the local approach one has to explicitly perform the back propagation of all the functions involved during the gate operations and computing the expectation values, which contains functions such as $\SVD$ whose back propagation could be very unstable. In addition, one could no longer use the memory efficient algorithm and all the intermediate calculations need to be saved in principle, which makes the local approach impractical.

\section{Results}\label{sec:result}

\begin{figure}[h]
  \centering
    \includegraphics[width=\linewidth]{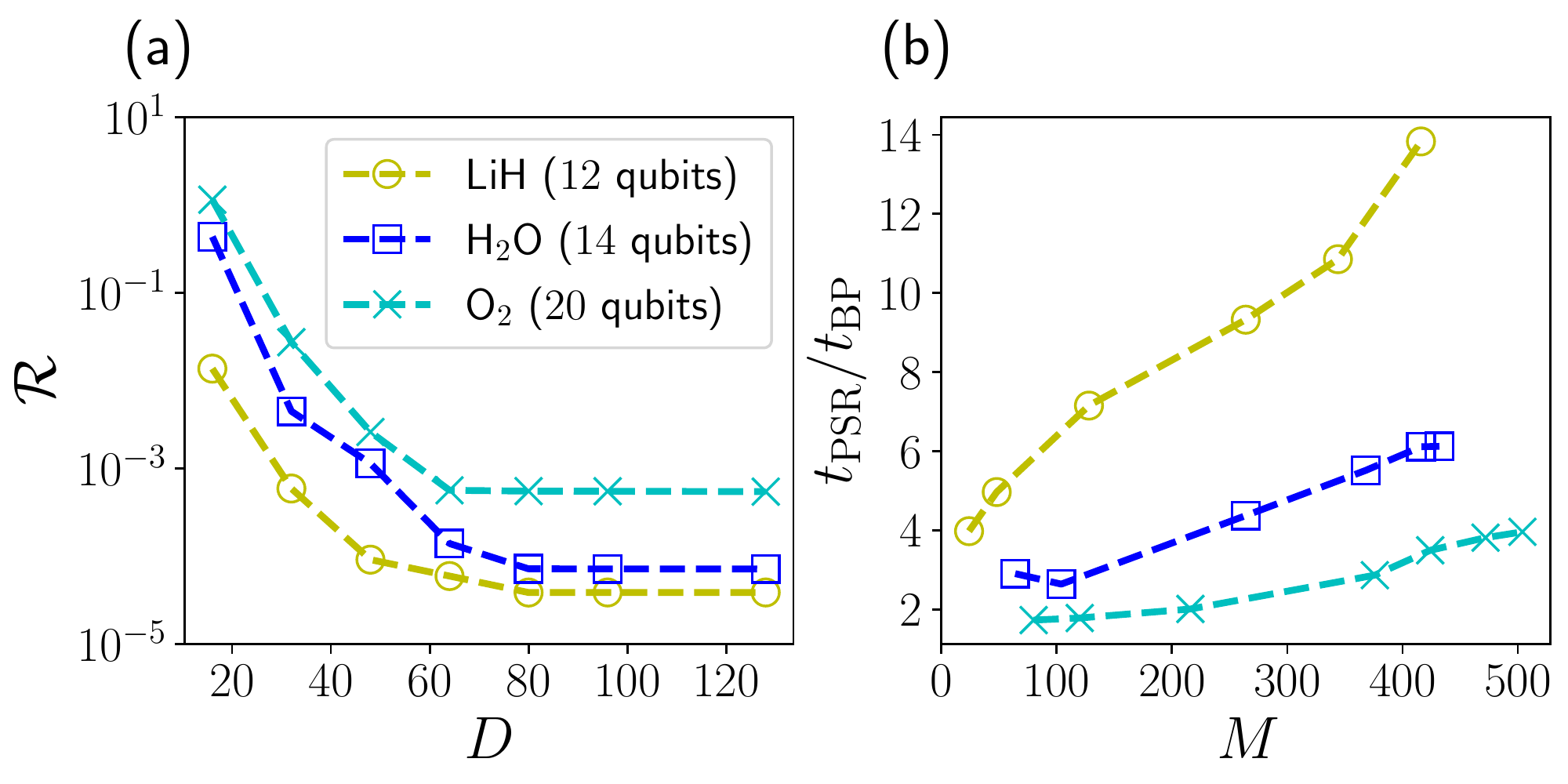}
    \caption{(a) The mean relative error $\rerror$, defined in Eq.(\ref{eq:meanerror}), as a function of the bond dimension $D$ for the LiH molecule (yellow dashed line with circle), the H$_2$O molecule (blue dashed line with square) and the O$_2$ molecule (cyan dashed line with $\times$), where the number of parameters used for these three molecules are $M=416,432,504$ respectively. (b) The speed up of the gradient calculation using differentiable MPS against using the parameter shift rule for the LiH, H$_2$O and O$_2$ molecules with a fixed $D=128$. In these simulations we have used $N_h /m = 8$. \gcc{Each point is calculated by averaging over $10$ independent evaluations with randomly initialized variational parameters within range $[-\pi, \pi]$.}
    }
    \label{fig:benchmarkgradients}
\end{figure}

In the following we first validate our differentiable MPS simulator by benchmarking it against PSR. Concretely we consider three molecules LiH, H$_2$O and O$_2$, and study the accuracy and the runtime efficiency of our differentiable MPS compared to the PSR. The results are shown in Fig.~\ref{fig:benchmarkgradients}.
In Fig.~\ref{fig:benchmarkgradients}(a) we study the mean relative error $\rerror$ between the gradients computed using differentiable MPS, denoted as $\gradbp $, and the gradients computed using Eq.(\ref{eq:psr}), denoted as $\gradpsr$, which is defined as
% \begin{align}
% \rerror = \frac{1}{M} \sum_{i=1}^M \left|\frac{\gradbp_i - \gradpsr_i}{\gradpsr_i}\right|.
% \end{align}
\begin{align}\label{eq:meanerror}
\rerror =  \frac{\Vert\gradbp - \gradpsr\Vert}{\Vert\gradpsr\Vert},
\end{align}
with $\Vert \bold{x} \Vert$ the Euclidean norm of the vector $\bold{x}$.
We can see that the gradients calculated by these two methods converge to each other as $D$ grows, as expected (we note that both $\gradpsr$ and $\gradbp $ are not exact if $D$ is not large enough, however the $\gradpsr$ is expected to be more precise than $\gradbp $, since in computing $\gradbp$ one needs in addition to apply the summation of $N_h/m$ Pauli strings onto the quantum state and we will generally use $m < N_h$). In Fig.~\ref{fig:benchmarkgradients}(b), we show the speed up of our differentiable MPS compared to PSR, evaluated by $t_{{\rm PSR}} / t_{{\rm BP}}$, where $t_{{\rm PSR} / {\rm BP}} $ are the runtimes for computing the gradients of a single optimization step, and \gcc{$t_{{\rm PSR}}$} is estimated by the runtime of the forward propagation times $2M$. We can see that the speed up scales approximately linearly as $M$ increases. 
% differentiable MPS becomes more advantageous than PSR as the number of parameters in the variational quantum circuit ansatz increases.

Here we note that in the variational quantum circuit ansatz used in VQE, there could often be several parametric quantum gates (more concretely the $R_z$ gate) which share the same parameter, and thus the actual number of parameters will be less than the number of parametric quantum gates. Nevertheless, in this work when we mention the number of parameters, we always mean the total number of parametric quantum gates, since 1) those parameters can be optimized independently in principle and 2) the complexities of both our differentiable MPS and the PSR are only determined by the latter. 

\begin{figure}[h]
  \centering
    \includegraphics[width=\linewidth]{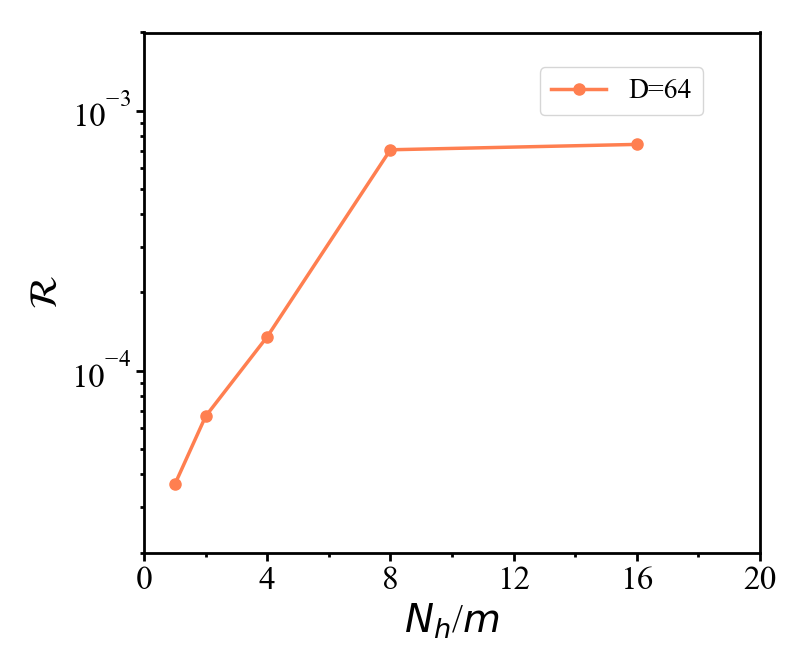}
    \caption{ \gcc{The mean relative error $\rerror$ against the number of Pauli strings in each group, namely $N_h/m$. We have used the H$_2$O molecule in the STO-3G basis set ($14$ qubits) for this calculation, for which $N_h=384$.}
    }
    \label{fig:gradientsvsm}
\end{figure}

\gcc{Now we study the effect of $m$ on the accuracy of the obtained gradient. As discussed in Sec.~\ref{sec:ad}, for a given Hamiltonian, we expect higher accuracy and computational cost for larger $m$ and vice versa. This effect is shown in Fig.~\ref{fig:gradientsvsm}. We can clearly see that the mean relative error $\rerror$ goes up when increasing the number of Pauli strings in each group (we note that the gradient calculated by PSR may also not be exact, but is expected to be more accurate than that calculated by our back propagation algorithm if $D$ is not large enough), and stabilizes to around $10^{-3}$ for $N_h/m \geq 8$ ($m \leq 48$).}

% In the folowing we first study the performance of the back propagation algorithm for our MPS-VQE.
% % in terms of the accuracy of the gradients and the runtime efficiency compared to the parameter shift rule. 
% Concretely, we take the LiH molecule with XXX basis ($12$ qubits) amnd the H$_2$ molecule with the Aug-cc-pVDZ basis ($36$ qubits) as examples,
% and study the accuracy of the gradients computed using our back propagation algorithm and the runtime efficiency compared to the parameter shift rule using Eq.(\ref{eq:psr}), and the results are shown in Fig.~\ref{fig:benchmarkgradients}. Now we denoted the gradients computed using back propagation as $\bold{g}^{bp} $

%\gcc{In the following we further check the accuracy of our VQE-MPS against the two most important hyperparameters, the bond dimension $D$ and the truncation threshold, denoted as $\epsilon$, which is shown in Fig.~\ref{fig:MPS-test}. }

\begin{figure}
\centering
  \includegraphics[width=\linewidth]{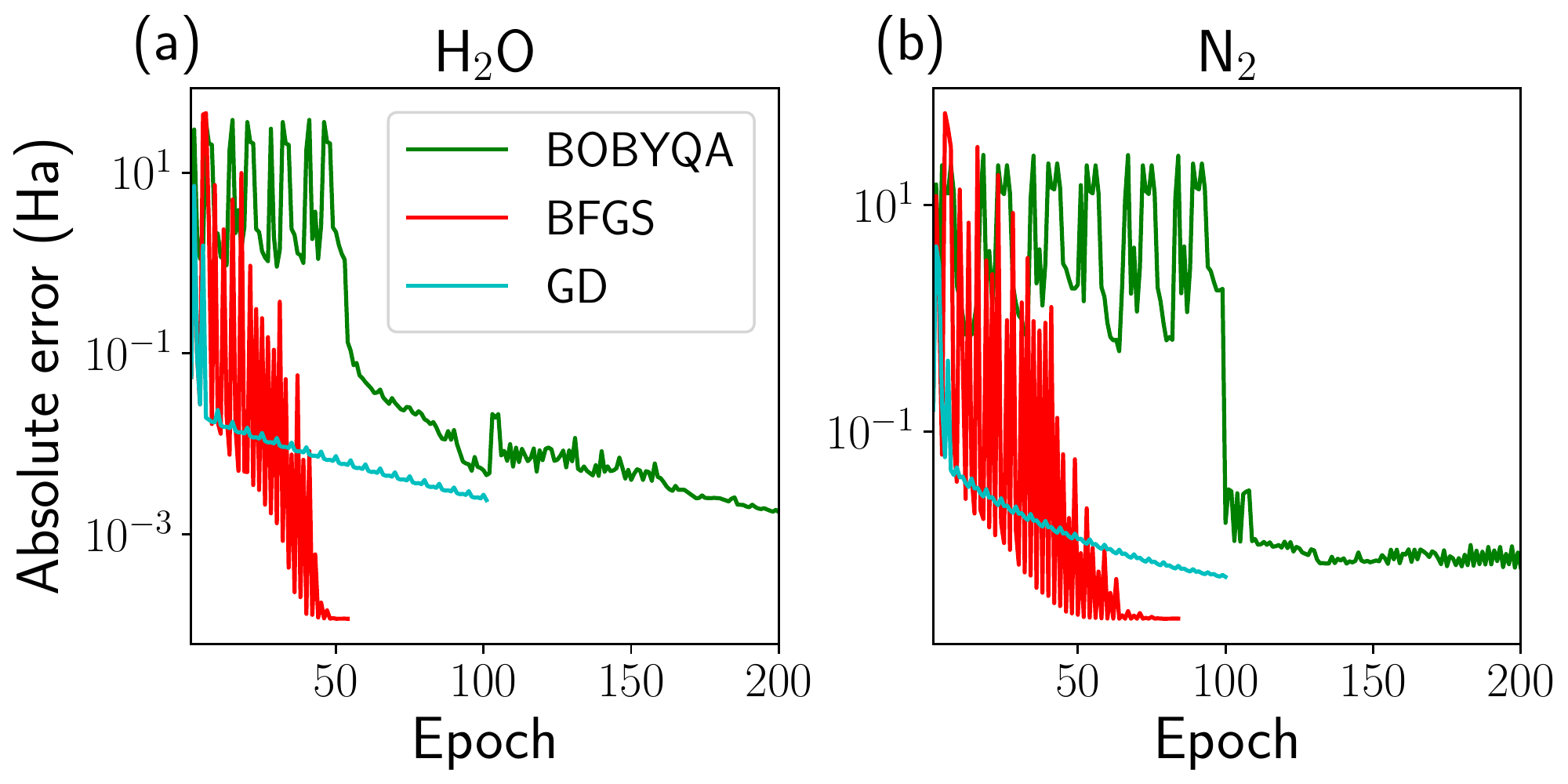}
  \caption{Convergence of different optimization algorithms towards the ground state for (a) the H$_2$O molecule and (b) the N$_2$ molecule. The green, red and cyan lines in both panels stand for the BOBYQA, BFGS and gradient descent (GD) algorithms respectively. The y axis show the absolute error between the energies computed using these methods and the FCI energy. In these simulations we have used $N_h / m = 4$. \gcc{The typical hyperparameters for these optimization algorithms are set as follows (which are used throughout this work). For gradient-free optimizer (BOBYQA) the convergence criterion is that the difference between two successive minimization steps (epochs) is smaller than $10^{-6}$. For gradient-based optimizers (GD and BFGS), we use previous criterion plus an additional criterion that the norm of the gradient is less than $10^{-5}$, and the program stops if any of the two criteria is met. The maximum number of iterations is set to be $100$ for gradient-based optimizers and $200$ for BOBYQA.}}
  \label{fig:convergence_H2O}
\end{figure}

% Then we compare our gradient-based optimization methods such as BFGS and Gradient Descent to classic gradient-free method such as BOBYQA. The details of three molecules we used in the test are shown in Table.~\ref{convergence}.

In VQE, there could be various reasons if the algorithm ends in a sub-optimal solution, such as 1) the poor expressiveness of the quantum circuit ansatz, 2) the inaccuracy during computing the gradients, and 3) the likelihood of being trapped in an excited state (for example the barren plateaus~\cite{McCleanNeven2018}). A usual approach to overcome the last pitfall is to start from a knowledged initial state (such as the Hartree-Fock reference state used here), while a (numerically) exact gradient could help us to eliminate the second one. Based on the exact gradient, it is also possible to use a very high-level optimization solver, which could be very helpful for one to examine the expressiveness of the underlying quantum circuit ansatz. In Fig.~\ref{fig:convergence_H2O}, we study the convergence of VQE for the H$_2$O and N$_2$ molecules using different optimization solvers, including the gradient-free solver BOBYQA, the gradient descent algorithm and the high-level BFGS algorithm~\cite{Powell_BOBYQA_2009, Jorge_OPT_1999}. We can see that for both molecules the gradient-based solvers converge much faster than BOBYQA, in particular the BFGS algorithm could converge to the lowest energy within less than $50$ epochs. Moreover, the gradient-free solver BOBYQA could still have high fluctuations beyond $100$ epochs when the energy is quite close to the exact (FCI) energy. In these simulations as well as in the later simulations we have always used a bond dimension $D=128$ and an SVD truncation threshold $\epsilon=10^{-6}$. For reference, the detailed information of the molecules and the corresponding variational ansatz used in Fig.~\ref{fig:convergence_H2O} as well as in our later simulations are listed in TABLE.~\ref{tab:info}. 

\begin{table*}[!htb]
  \begin{center}
    \caption{Information of the molecules used in this work. The columns ``Qubits'', ``Parameters'', ``Pauli strings'', ``Basis set'' list the number of qubits, the number of parametric gates (after being reduced by symmetry), the number of Pauli strings, and the basis set we used for the molecule.}
    \label{tab:info}
    \begin{tabular}{|c|c|c|c|c|c}
    \hline 
      Molecule & Qubits & Parameters & Pauli strings & Basis set\\
\hline
LiH & 12 & 416 & 227 & STO-3G\\
H$_2$O & 14 & 432 & 384 & STO-3G\\
N$_2$ & 20 & 928 & 749 & STO-3G\\
% H$_2$ & 36 & 41 & 1290 & Aug-cc-pVDZ\\
O$_2$ & 20 & 504 & 740 & STO-3G\\
HF & 12 & 100 & 227 & STO-3G\\
HCl & 20 & 284 & 1303 & STO-3G\\
\hline
    \end{tabular}
  \end{center}
\end{table*}

% \subsection{Validation against benchmark results}

\begin{figure*}
\centering
  \includegraphics[width=\textwidth]{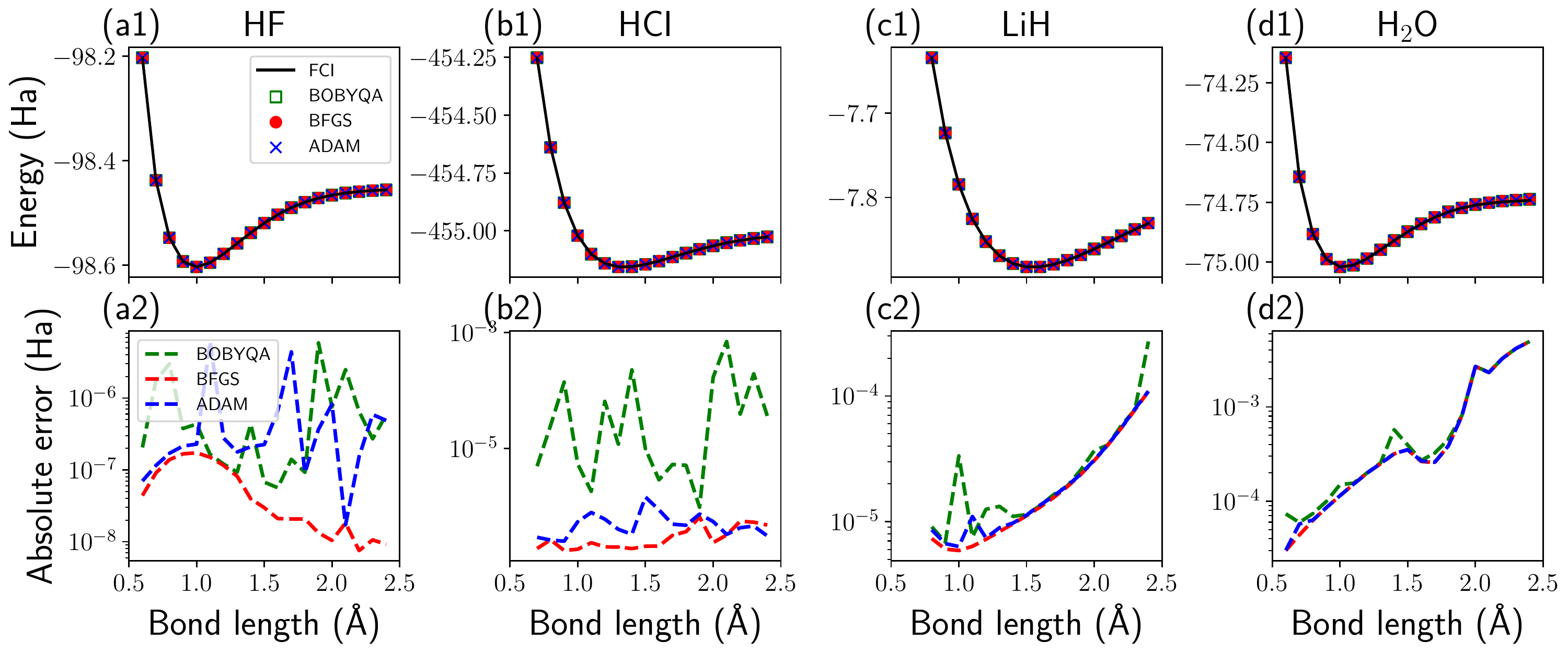}
  \caption{The potential energy curve for four molecules HF (a1), HCl (b1), LiH (c1) and H$_2$O (d1) using the STO-3G basis set. In (a1,b1,c1,d1) the green squares represent the results obtained by the gradient-free BOBYQA optimizer, the red circles are the results obtained by the BFGS optimizer, the blue $\times$ are the results obtained by the ADAM optimizer with an initial learning rate $0.01$, and the black solid lines represent the FCI energies for comparison. (a2,b2,c2,d2) The absolute errors between the results obtained by the three optimizers BOBYQA, BFGS and ADAM compared to the corresponding FCI energies for HF (a2), HCl (b2), LiH (c2) and H$_2$O (d2). In these simulations we have used $N_h / m = 2$. 
  % We have set the tolerance for the absolute value of the loss function to be $10^{-8}$ as the stop criterion for all the optimizers. For the BFGS algorithm we have set the tolerance for the absolute norm of the gradient to be $10^{-5}$ as an additional stop criterion.
  }
  \label{fig:correct}
\end{figure*}

To demonstrate the powerfulness of our differentiable MPS simulator for large molecules, we compute the potential energy curves for the HF, HCl, LiH, H$_2$O molecules. 
% For these simulations we have used the BFGS optimization solver with $800$ iterations for the HF, HCl and LiH molecules, and with $1200$ iterations for the H$_2$O molecule (for which the number of parameters is larger than others) to ensure convergence in all cases (\gcc{Do we really need such a large number of iterations even with BFGS?}). We have also used a bond dimension $D=128$ and SVD truncation threshold $\epsilon=10^{-5}$ in these simulations, and the results obtained using BOBYQA and BFGS are shown as a comparison.
% We can see that for the HF, HCl and LiH molecules and for bond length ranging from $0.5$ \r{A} to $2.5$ \r{A}, both BOBYQA and BFGS result in accurate energies with absolute errors compared to the FCI energies less than $10^{-3}$ Ha (chemical precision). For bond length larger than $2.5$ \r{A}, we could not obtain reasonable results using the BOBYQA optimizer except for LiH. Nevertheless, with BFGS we can still obtain reasonable energies in this regime for all the considered molecules. In particular, the energies obtained with BFGS for bond length larger than $2.5$ \r{A} still have absolute errors less than $10^{-5}$. 
Symmetry-reduced UCCSD\cite{CaoYung2022} is used as the variational ansatz, leading to 100$\sim$432 Rz gates as shown in TABLE.~\ref{tab:info} for the testing molecules.
The results are compared with the FCI energies as shown in Fig.~\ref{fig:correct}. We can see that for all the molecules with bond length ranging from $0.5$ \r{A} to $2.0$ \r{A}, chemical accuracy (error within $1.6\times10^{-3}$ Hatree or 1 kcal/mol) is reached. Although all the optimization algorithms give accurate energies, the gradient-based Adam and BFGS algorithms can achieve smaller absolute errors than the gradient-free BOBYQA, especially for HF and HCl. Generally, BFGS outperforms other method and achieves the best accuracy.
\begin{table*}[!htb]
  \begin{center}
    \caption{Ground state energies for the molecules O$_2$, CO, CO$_2$ involved in the chemical reaction in Eq.(\ref{eq:co2}). Units are in Hartree. We have used $N_h /m = 8$ in this simulation.
    }
    \label{tab:co2}
    \begin{tabular}{|c|c|c|c|}
    \hline
     & O$_2$ & CO & CO$_2$ \\
    \hline
Basis set & STO-3G & STO-3G & STO-3G \\
Qubits & 20 & 20 & 30 \\
UCCSD operators & 19 & 60 & 55 \\
Pauli strings & 2239 & 4427 & 11430 \\
% HF & -147.63216204  &  -111.22458955 & -185.06469568 \\
CCSD & -147.7419 & -111.3555 & -185.2559 \\
% CCSD(T) & -147.74259802 & -111.36256738 & -185.27645022 \\
FCI & -147.7440 & -111.3634 & -185.2761 \\
VQE results & -147.7425 & -111.3527 & -185.2574 \\
% BFGS results & \fanyi{-147.69664057} & \fanyi{-111.33415291} & \fanyi{-185.18102706} \\
%\hline
%$\Delta E$ & \multicolumn{3}{c|}{-18.17979769} \\
\hline
    \end{tabular}
  \end{center}
\end{table*}

To this end we note that till now we have only studied relatively small molecules with no more than $20$ qubits (TABLE.~\ref{tab:info}). However, we stress that the major limitations of our method is the entanglement entropy within the quantum state (which determines the minimal bond dimension require to maintain certain precision), instead of the number of qubits, the circuit depth as well as the number of parameters of the underlying quantum circuit ansatz. The largest VQE simulation using the state-vector based simulators till now have computed the ground state energy of the C$_2$H$_4$ molecule with $28$ qubits~\cite{CaoYung2022}. In the following we go a step forward by computing the energies of the CO, O$_2$ and CO$_2$ molecules involved in the chemical reaction:
\begin{align}\label{eq:co2}
    CO + \frac{1}{2}O_2 \longrightarrow CO_2 ,
\end{align}
%Ansatz of CO, O2 and CO2 molecules are generated by symmetry reduced UCCSD and MP2.
and the results are shown in TABLE.~\ref{tab:co2}. Previous studies have calculated this reaction using a state vector simulator by incorporating the frozen core approximation and qubit tapering~\cite{Sapova_CO2_2022}. Here we performed the full-size simulation of this 30-qubit system. For demonstration purpose, we selected operators using the point group symmetry~\cite{CaoYung2022} and energy-sorting screening~\cite{ES_2021}. The obtained energy barrier is $-21.0$ kcal/mol. Compared with FCI ($-25.5$ kcal/mol), there is an error of $4.5$ kcal/mol, mainly caused by the incompleteness of the screened UCCSD operator pool (especially for CO$_2$). More accurate results can be obtained if more robust ansatz such as UCC with generalized excitations (UCCGSD)~\cite{uccgsd} or ADAPT-VQE~\cite{ADAPT-VQE} are used. 

\begin{figure}
\centering
  \includegraphics[width=\columnwidth]{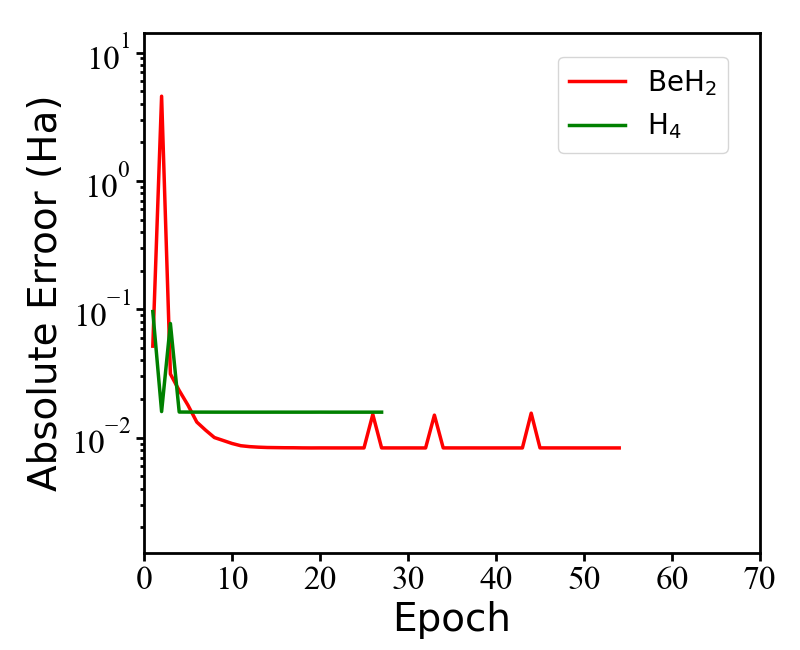}
  \caption{\gcc{Convergence of the energy during VQE for the BeH$_2$ molecule in the 6-31G* basis set and the H$_4$ molecule in the cc-pVDZ basis set, where we have used the BFGS optimizer. The y-axis shows the absolute error compared to the FCI energy. The detailed information for these simulations are listed in Table.~\ref{tab:large}.}
  }
  \label{fig:BeH2}
\end{figure}

\gcc{To further demonstrate the powerfulness of our method, we study the BeH$_2$ molecule in the 6-31G* basis set and H$_4$ molecule in the cc-pVDZ basis set, with $36$ and $40$ qubits respectively. Geometry optimization is first performed using PySCF~\cite{pyscf}, and the equilibrium geometries are used to perform VQE calculations. The convergence of these two simulations against the minimization step is shown in Fig.~\ref{fig:BeH2}. In the meantime, we list in Table.~\ref{tab:large} the complete information, including the actual memory and time costs, of our simulation for these two molecules plus CO$_2$ (the runtime per epoch for BeH$_2$ is the largest since its variational circuit ansatz contains the largest number of gate operations). We can see that in both cases the energy quickly converges to a steady value that is about $10^{-2}$ Hatree higher than the corresponding FCI energy, that is, one order of magnitude larger the chemical accuracy. Similar to the case of CO$_2$, the reason is likely due to the limited expressiveness of the wave function ansatz we have used. Since we have repeated each simulation for a large number of times with random initialization of the variational parameters and picked the one with the lowest final energy, we believe that these simulations are less likely trapped in local minima. The relatively stable convergence indicates that the gradients calculated with our differentiable MPS should be fairly accurate, while exploring the convergence of large molecules towards chemical accuracy using more involved variational circuit ansatz is beyond the scope of this work.}

\begin{table*}[!htb]
  \begin{center}
    \caption{ \gcc{Information of our simulations for the three molecules CO$_2$ (STO-3G), BeH$_2$ (6-31G*) and H$_4$ (cc-pVDZ) with equal to or more than $30$ qubits. We have used $N_h/m=4$ and $70$ processes for all these simulations. Each process uses $8$ threads on an Intel x86 CPU with $2.7$ GHz frequency per thread. The column ``Memory'' shows the largest memory usage per process and the column ``Time'' shows the runtime per epoch averaged over $5$ epochs.}
    }
    \label{tab:large}
    \begin{tabular}{|c|c|c|c|c|c|c|}
    \hline
    Molecule & $N$ & $N_h$ & $D$ & $N_g$ & Memory (GB) & Time (s) \\
    \hline
CO$_2$ & $30$ & $3232$ & $128$ & $45534$ & $19.71$ & $3978$  \\
BeH$_2$ & $36$ & $5553$ & $128$ & $59234$ & $19.43$ & $14909$ \\
H$_2$ & $40$ & $9516$ & $96$ & $46124$ & $23.43$ & $7233$ \\
\hline
    \end{tabular}
  \end{center}
\end{table*}

\section{Conclusions}
\label{sec:conclusions}
% In this work, we propose a massively parallel quantum simulation method based on MPI.jl, combined with the memory-efficient scheme accelerating the circuit measurement and VQE iteration. In HF, HCl and LiH molecule systems, the absolute error of VQE simulation is under $10^-3$ Ha. In H$_2$O molecule, the accuracy requirement is meet in short bond length. Besides, the strong and weak scalability of parallel method is excellent up to 2048 processes, with memory-efficient scheme proven to be effective.

In this work, we have proposed the differentiable matrix product states for simulating variational quantum algorithms. The differentiable MPS simulator seamlessly integrates the MPS based quantum circuit simulator with the automatic differentiation framework, by using a memory-efficient back propagation algorithm where only two copies of quantum states have to be stored at least. Compared to other approaches for simulating variational quantum circuits, our differentiable MPS is less sensitive to the number of qubits, and its complexity is independent of the number of variable parameters. The accuracy as well as the efficiency of our method are demonstrated in real chemical systems, which shows that our differentiable MPS simulator could be a promising testing ground for variational quantum algorithms on near-term quantum computers.

\gcc{In the end, we stress that as a general feature of MPS based algorithms, the bond dimension of the MPS ansatz could significantly affect the accuracy of the simulation, depending on the entanglement generated in the underlying quantum state. For our differentiable MPS simulator which aims to simulate variational quantum circuits, the calculated gradients also subject to this constrain. Therefore a good practice in applying the differentiable MPS simulator is to check the convergence of it against different bond dimensions. In future investigations, other types of tensor networks, such as the projected entangled pair states~\cite{GuoWu2019} and the tree tensor network~\cite{HikiharaNishino2023}, could also be explored to simulate variational quantum circuits.}

\section{Acknowledgments}
\gcc{Our code for the differentiable MPS algorithm is purely written in the Julia language~\cite{Bezanson_JULIA_2017}, which works on any CPU that supports Julia. The parallelization is done using the MPI.jl package~\cite{byrne2021mpi}. The serial version of our code is open sourced~\cite{diffMPS}, which supports the splitting of the Hamiltonian buts runs in serial for different groups.}
\gcc{The optimization algorithms used in this work are from the NLopt.jl package~\cite{Steven_NLopt_2007}.}
% The code for our differentiable MPS algorithm is written in pure Julia language~\cite{XXX}, which can be found at~\cite{diffMPS}.}
This work was supported by National Natural Science Foundation of China under Grant No.~T2222026 and Grant No.~22003073. 
C. G. is supported by the Open Research Fund from State Key Laboratory of High Performance Computing of China (Grant No. 202201-00).

% \section{Data Availability}

% The data that supports the findings of this study are available within the article and its Appendix.

\appendix
%\section{Current ADAPT-VQE results}
%\begin{table}[H]
%  \begin{center}
%    \caption{C$O_2$ with 30 qubits, 167 parameters and 3232 hamiltonian terms. %FCI=-185.2761249497911251 Ha. Running in 29 nodes and 56 cores per node. %Adapt\_g\_tol=1e-3, adapt\_f\_tol=1e-5, D=50, $\epsilon$=1e-9, vqe\_f\_tol=1e-6, %vqe\_g\_tol=1e-4. Add 1 operator per iteration.}
%    \begin{tabular}{c|c|c}
%      Iteration & Measure time per operator & Optimized energy after vqe \\
%\hline
%1 & 0.066399 & -185.08776385134766 \\
%2 & 0.107174 & -185.11062805169968 \\
%3 & 0.174401 & -185.1229641389258 \\
%4 & 1.126862 & -185.12668476705196 \\
%5 & 12.15569 & -185.1300565784314 \\
%6 & 17.84636 & -185.1341827631172 \\
%7 & 20.24650 & -185.13846924784553 \\
%8 & 20.25427 & -185.13910420939257 \\
%    \end{tabular}
%  \end{center}
%\end{table}

\section{Back propagation rule for loss function with complex numbers}\label{app:bprules}
Here we briefly summarize the back propagation rules defined in Ref.~\cite{GuoPoletti2021} for loss functions involving complex numbers (the output is still real). 
We assume that the loss function can be written as $F(z, z^{\ast})$ with $z = x + \im y$ and $z^{\ast} = x-\im y$. For simplicity we assume $x$, $y$ to be real numbers, but the results can be easily generalized to arrays of scalars. $F(z, z^{\ast})$ can also be equivalently viewed as a function of real numbers as $F(z, z^{\ast}) = \bar{F}(x, y)$, and we have the relations
\begin{align}
\frac{\partial F}{\partial z} &= \frac{1}{2}\left( \frac{\partial \bar{F}}{\partial x} -\im \frac{\partial \bar{F}}{\partial y}  \right); \\
\frac{\partial F}{\partial z^{\ast}} &= \frac{1}{2}\left( \frac{\partial \bar{F}}{\partial x} +\im \frac{\partial \bar{F}}{\partial y}  \right).
\end{align}
If we treat the loss function as a function of real numbers, it is well-known that the gradient of $\bar{F}$ should be $(-\frac{\partial \bar{F}}{\partial x},-\frac{\partial \bar{F}}{\partial y} )$ since we have
\begin{align}
&\bar{F}(x - \lambda \frac{\partial \bar{F}}{\partial x}, y -  \frac{\partial \bar{F}}{\partial y}) - \bar{F}(x, y) \nonumber \\ 
=& -\lambda \left[\left(\frac{\partial \bar{F}}{\partial x}\right)^2 + \left(\frac{\partial \bar{F}}{\partial y}\right)^2  \right] + O(\lambda^2),
\end{align}
where $\lambda$ is a small positive number. If we treat the loss function as a function of complex numbers, we can verify that if we update $z$ as $z - 2\lambda \frac{\partial F}{\partial z^{\ast}}$, then we have
\begin{align}
& F(z - 2\lambda \frac{\partial F}{\partial z^{\ast}}, z^{\ast} - 2\lambda \frac{\partial F}{\partial z} ) - F(z, z^{\ast})  \nonumber \\
=& -4\lambda \frac{\partial F}{\partial z} \frac{\partial F}{\partial z^{\ast}} + O(\lambda^2) \nonumber \\ 
=& -\lambda \left[\left(\frac{\partial \bar{F}}{\partial x}\right)^2 + \left(\frac{\partial \bar{F}}{\partial y}\right)^2  \right] + O(\lambda^2),
\end{align}
which means that we can simply take $-2 \frac{\partial F}{\partial z^{\ast}}$ as the gradient of $F(z, z^{\ast})$ against $z$ when minimizing $F(z, z^{\ast})$.

Now we assume that $F(z, z^{\ast})$ can be written as a composite function as
\begin{align}
F(z, z^{\ast}) = f \circ g \circ h \circ \cdots \circ p (z, z^{\ast}).
\end{align}
It is proven in Ref.~\cite{GuoPoletti2021} that if we define the back propagation rule for each elementary function as
\begin{align}\label{eq:bp1}
\tilde{g}(\nu^{\ast})|_z = \nu \frac{\partial g}{\partial z^{\ast}} + \nu^{\ast} \frac{\partial g^{\ast}}{\partial z^{\ast}},
\end{align}
then the gradient of $F(z, z^{\ast})$ against $z$ can be evaluated as
\begin{align}
2 \frac{\partial F}{\partial z^{\ast}} = \tilde{p}\circ \cdots \circ \tilde{h}\circ \tilde{g}\circ \tilde{f}(1),
\end{align}
which is exactly in parallel with the case of real loss functions.
For nonscalar complex function $g$, one can simply generalize Eq.(\ref{eq:bp1}) as
\begin{align}\label{eq:bp2}
\tilde{g}_j(\nu^{\ast})|_z = \sum_i \nu_i \frac{\partial g_i}{\partial z^{\ast}_j} + \nu^{\ast}_i \frac{\partial g_i^{\ast}}{\partial z_j^{\ast}}.
\end{align}
In case the input of $g$ is real, Eq.(\ref{eq:bp2}) reduces to 
\begin{align}\label{eq:bp3}
\tilde{g}_j(\nu^{\ast})|_z = 2 {\rm Re}\left( \sum_i\nu_i \frac{\partial g_i}{\partial z^{\ast}_j} \right).
\end{align}
In case the output of $g$ is real, Eq.(\ref{eq:bp2}) reduces to 
\begin{align}\label{eq:bp4}
\tilde{g}_j(\nu^{\ast})|_z = 2\sum_i \left({\rm Re}(\nu_i)\frac{\partial g_i}{\partial z^{\ast}_j}\right).
\end{align}
Eq.(\ref{eq:bp3}) and Eq.(\ref{eq:bp4}) are the back propagation rules used to derive Eq.(\ref{eq:circuitevolvegrad}) and Eq.(\ref{eq:expecgrad}) in the main text respectively.

%%%%%%%%%%%%%%%%%%%%%%%%%%%%%%%%%%%%%%%%%%%%%%%%%%%%%%%%%%%%%%%%%%
%                        Bibliography                            %
%%%%%%%%%%%%%%%%%%%%%%%%%%%%%%%%%%%%%%%%%%%%%%%%%%%%%%%%%%%%%%%%%%
%\bibliographystyle{apsrev4-1}
%\bibliographystyle{plainnat}
\bibliography{refs_submit_final} % Produces the bibliography via BibTeX.

\begin{thebibliography}{66}
\providecommand{\natexlab}[1]{#1}
\providecommand{\url}[1]{\texttt{#1}}
\expandafter\ifx\csname urlstyle\endcsname\relax
  \providecommand{\doi}[1]{doi: #1}\else
  \providecommand{\doi}{doi: \begingroup \urlstyle{rm}\Url}\fi

\bibitem[Arute et~al.(2019)Arute, Arya, Babbush, Bacon, Bardin, Barends,
  Biswas, Boixo, Brandao, Buell, Burkett, Chen, Chen, Chiaro, Collins,
  Courtney, Dunsworth, Farhi, Foxen, Fowler, Gidney, Giustina, Graff, Guerin,
  Habegger, Harrigan, Hartmann, Ho, Hoffmann, Huang, Humble, Isakov, Jeffrey,
  Jiang, Kafri, Kechedzhi, Kelly, Klimov, Knysh, Korotkov, Kostritsa, Landhuis,
  Lindmark, Lucero, Lyakh, Mandrà, McClean, McEwen, Megrant, Mi, Michielsen,
  Mohseni, Mutus, Naaman, Neeley, Neill, Niu, Ostby, Petukhov, Platt, Quintana,
  Rieffel, Roushan, Rubin, Sank, Satzinger, Smelyanskiy, Sung, Trevithick,
  Vainsencher, Villalonga, White, Yao, Yeh, Zalcman, Neven, and
  Martinis]{AruteMartinis2019}
Frank Arute, Kunal Arya, Ryan Babbush, Dave Bacon, Joseph~C. Bardin, Rami
  Barends, Rupak Biswas, Sergio Boixo, Fernando G. S.~L. Brandao, David~A.
  Buell, Brian Burkett, Yu~Chen, Zijun Chen, Ben Chiaro, Roberto Collins,
  William Courtney, Andrew Dunsworth, Edward Farhi, Brooks Foxen, Austin
  Fowler, Craig Gidney, Marissa Giustina, Rob Graff, Keith Guerin, Steve
  Habegger, Matthew~P. Harrigan, Michael~J. Hartmann, Alan Ho, Markus Hoffmann,
  Trent Huang, Travis~S. Humble, Sergei~V. Isakov, Evan Jeffrey, Zhang Jiang,
  Dvir Kafri, Kostyantyn Kechedzhi, Julian Kelly, Paul~V. Klimov, Sergey Knysh,
  Alexander Korotkov, Fedor Kostritsa, David Landhuis, Mike Lindmark, Erik
  Lucero, Dmitry Lyakh, Salvatore Mandrà, Jarrod~R. McClean, Matthew McEwen,
  Anthony Megrant, Xiao Mi, Kristel Michielsen, Masoud Mohseni, Josh Mutus,
  Ofer Naaman, Matthew Neeley, Charles Neill, Murphy~Yuezhen Niu, Eric Ostby,
  Andre Petukhov, John~C. Platt, Chris Quintana, Eleanor~G. Rieffel, Pedram
  Roushan, Nicholas~C. Rubin, Daniel Sank, Kevin~J. Satzinger, Vadim
  Smelyanskiy, Kevin~J. Sung, Matthew~D. Trevithick, Amit Vainsencher, Benjamin
  Villalonga, Theodore White, Z.~Jamie Yao, Ping Yeh, Adam Zalcman, Hartmut
  Neven, and John~M. Martinis.
\newblock Quantum supremacy using a programmable superconducting processor.
\newblock \emph{Nature}, 574\penalty0 (7779):\penalty0 505--510, 2019.
\newblock \doi{doi.org/10.1038/s41586-019-1666-5}.

\bibitem[Wu et~al.(2021)Wu, Bao, Cao, Chen, Chen, Chen, Chung, Deng, Du, Fan,
  Gong, Guo, Guo, Guo, Han, Hong, Huang, Huo, Li, Li, Li, Li, Liang, Lin, Lin,
  Qian, Qiao, Rong, Su, Sun, Wang, Wang, Wu, Xu, Yan, Yang, Yang, Ye, Yin,
  Ying, Yu, Zha, Zhang, Zhang, Zhang, Zhang, Zhao, Zhao, Zhou, Zhu, Lu, Peng,
  Zhu, and Pan]{WuPan2021}
Yulin Wu, Wan-Su Bao, Sirui Cao, Fusheng Chen, Ming-Cheng Chen, Xiawei Chen,
  Tung-Hsun Chung, Hui Deng, Yajie Du, Daojin Fan, Ming Gong, Cheng Guo, Chu
  Guo, Shaojun Guo, Lianchen Han, Linyin Hong, He-Liang Huang, Yong-Heng Huo,
  Liping Li, Na~Li, Shaowei Li, Yuan Li, Futian Liang, Chun Lin, Jin Lin,
  Haoran Qian, Dan Qiao, Hao Rong, Hong Su, Lihua Sun, Liangyuan Wang, Shiyu
  Wang, Dachao Wu, Yu~Xu, Kai Yan, Weifeng Yang, Yang Yang, Yangsen Ye,
  Jianghan Yin, Chong Ying, Jiale Yu, Chen Zha, Cha Zhang, Haibin Zhang, Kaili
  Zhang, Yiming Zhang, Han Zhao, Youwei Zhao, Liang Zhou, Qingling Zhu,
  Chao-Yang Lu, Cheng-Zhi Peng, Xiaobo Zhu, and Jian-Wei Pan.
\newblock Strong quantum computational advantage using a superconducting
  quantum processor.
\newblock \emph{Phys. Rev. Lett.}, 127:\penalty0 180501, Oct 2021.
\newblock \doi{10.1103/PhysRevLett.127.180501}.

\bibitem[Zhu et~al.(2022)Zhu, Cao, Chen, Chen, Chen, et~al.]{ZhuPan2022}
Qingling Zhu, Sirui Cao, Fusheng Chen, Ming-Cheng Chen, Xiawei Chen, et~al.
\newblock Quantum computational advantage via 60-qubit 24-cycle random circuit
  sampling.
\newblock \emph{Science Bulletin}, 67\penalty0 (3):\penalty0 240--245, 2022.
\newblock \doi{doi.org/10.1016/j.scib.2021.10.017}.

\bibitem[Wang et~al.(2019)Wang, Higgott, and Brierley]{AVQE}
Daochen Wang, Oscar Higgott, and Stephen Brierley.
\newblock Accelerated variational quantum eigensolver.
\newblock \emph{Phys. Rev. Lett.}, 122:\penalty0 140504, Apr 2019.
\newblock \doi{10.1103/PhysRevLett.122.140504}.

\bibitem[DiAdamo et~al.(2021)DiAdamo, Ghibaudi, and Cruise]{DVQE}
Stephen DiAdamo, Marco Ghibaudi, and James Cruise.
\newblock Distributed quantum computing and network control for accelerated
  vqe.
\newblock \emph{IEEE Transactions on Quantum Engineering}, 2:\penalty0 1–21,
  2021.
\newblock ISSN 2689-1808.
\newblock \doi{10.1109/tqe.2021.3057908}.

\bibitem[Lolur et~al.(2021)Lolur, Rahm, Skogh, García-Álvarez, and
  Wendin]{VQEbenchmark}
P.~Lolur, M.~Rahm, M.~Skogh, L.~García-Álvarez, and G.~Wendin.
\newblock Benchmarking the variational quantum eigensolver through simulation
  of the ground state energy of prebiotic molecules on high-performance
  computers.
\newblock \emph{AIP Conference Proceedings}, 2362\penalty0 (1):\penalty0
  030005, 2021.
\newblock \doi{10.1063/5.0054915}.

\bibitem[Cao et~al.(2019)Cao, Romero, Olson, Degroote, Johnson, Kieferov{\'a},
  Kivlichan, Menke, Peropadre, Sawaya, et~al.]{CaoGuzik2019}
Yudong Cao, Jonathan Romero, Jonathan~P Olson, Matthias Degroote, Peter~D
  Johnson, M{\'a}ria Kieferov{\'a}, Ian~D Kivlichan, Tim Menke, Borja
  Peropadre, Nicolas~PD Sawaya, et~al.
\newblock Quantum chemistry in the age of quantum computing.
\newblock \emph{Chemical reviews}, 119\penalty0 (19):\penalty0 10856--10915,
  2019.
\newblock \doi{doi.org/10.1021/acs.chemrev.8b00803}.

\bibitem[null null et~al.(2020)null null, Arute, Arya, Babbush, Bacon, Bardin,
  Barends, Boixo, Broughton, Buckley, Buell, Burkett, Bushnell, Chen, Chen,
  Chiaro, Collins, Courtney, Demura, Dunsworth, Farhi, Fowler, Foxen, Gidney,
  Giustina, Graff, Habegger, Harrigan, Ho, Hong, Huang, Huggins, Ioffe, Isakov,
  Jeffrey, Jiang, Jones, Kafri, Kechedzhi, Kelly, Kim, Klimov, Korotkov,
  Kostritsa, Landhuis, Laptev, Lindmark, Lucero, Martin, Martinis, McClean,
  McEwen, Megrant, Mi, Mohseni, Mruczkiewicz, Mutus, Naaman, Neeley, Neill,
  Neven, Niu, O’Brien, Ostby, Petukhov, Putterman, Quintana, Roushan, Rubin,
  Sank, Satzinger, Smelyanskiy, Strain, Sung, Szalay, Takeshita, Vainsencher,
  White, Wiebe, Yao, Yeh, and Zalcman]{Google2022a}
null null, Frank Arute, Kunal Arya, Ryan Babbush, Dave Bacon, Joseph~C. Bardin,
  Rami Barends, Sergio Boixo, Michael Broughton, Bob~B. Buckley, David~A.
  Buell, Brian Burkett, Nicholas Bushnell, Yu~Chen, Zijun Chen, Benjamin
  Chiaro, Roberto Collins, William Courtney, Sean Demura, Andrew Dunsworth,
  Edward Farhi, Austin Fowler, Brooks Foxen, Craig Gidney, Marissa Giustina,
  Rob Graff, Steve Habegger, Matthew~P. Harrigan, Alan Ho, Sabrina Hong, Trent
  Huang, William~J. Huggins, Lev Ioffe, Sergei~V. Isakov, Evan Jeffrey, Zhang
  Jiang, Cody Jones, Dvir Kafri, Kostyantyn Kechedzhi, Julian Kelly, Seon Kim,
  Paul~V. Klimov, Alexander Korotkov, Fedor Kostritsa, David Landhuis, Pavel
  Laptev, Mike Lindmark, Erik Lucero, Orion Martin, John~M. Martinis, Jarrod~R.
  McClean, Matt McEwen, Anthony Megrant, Xiao Mi, Masoud Mohseni, Wojciech
  Mruczkiewicz, Josh Mutus, Ofer Naaman, Matthew Neeley, Charles Neill, Hartmut
  Neven, Murphy~Yuezhen Niu, Thomas~E. O’Brien, Eric Ostby, Andre Petukhov,
  Harald Putterman, Chris Quintana, Pedram Roushan, Nicholas~C. Rubin, Daniel
  Sank, Kevin~J. Satzinger, Vadim Smelyanskiy, Doug Strain, Kevin~J. Sung,
  Marco Szalay, Tyler~Y. Takeshita, Amit Vainsencher, Theodore White, Nathan
  Wiebe, Z.~Jamie Yao, Ping Yeh, and Adam Zalcman.
\newblock Hartree-fock on a superconducting qubit quantum computer.
\newblock \emph{Science}, 369\penalty0 (6507):\penalty0 1084--1089, 2020.
\newblock \doi{10.1126/science.abb9811}.

\bibitem[Cheuk et~al.(2016)Cheuk, Nichols, Lawrence, Okan, Zhang, Khatami,
  Trivedi, Paiva, Rigol, and Zwierlein]{Google2022b}
Lawrence~W. Cheuk, Matthew~A. Nichols, Katherine~R. Lawrence, Melih Okan, Hao
  Zhang, Ehsan Khatami, Nandini Trivedi, Thereza Paiva, Marcos Rigol, and
  Martin~W. Zwierlein.
\newblock Observation of spatial charge and spin correlations in the 2d
  fermi-hubbard model.
\newblock \emph{Science}, 353\penalty0 (6305):\penalty0 1260--1264, 2016.
\newblock \doi{10.1126/science.aag3349}.

\bibitem[Schollwöck(2011)]{Schollwock2011}
Ulrich Schollwöck.
\newblock The density-matrix renormalization group in the age of matrix product
  states.
\newblock \emph{Annals of Physics}, 326\penalty0 (1):\penalty0 96–192, Jan
  2011.
\newblock ISSN 0003-4916.
\newblock \doi{10.1016/j.aop.2010.09.012}.

\bibitem[Or{\'u}s(2014)]{Orus2014}
Rom{\'a}n Or{\'u}s.
\newblock A practical introduction to tensor networks: Matrix product states
  and projected entangled pair states.
\newblock \emph{Annals of physics}, 349:\penalty0 117--158, 2014.
\newblock \doi{doi.org/10.1016/j.aop.2014.06.013}.

\bibitem[LaRose(2019)]{LaRose2019overviewcomparison}
Ryan LaRose.
\newblock Overview and {C}omparison of {G}ate {L}evel {Q}uantum {S}oftware
  {P}latforms.
\newblock \emph{{Quantum}}, 3:\penalty0 130, March 2019.
\newblock ISSN 2521-327X.
\newblock \doi{10.22331/q-2019-03-25-130}.

\bibitem[Jones et~al.(2019)Jones, Brown, Bush, and Benjamin]{Quest}
Tyson Jones, Anna Brown, Ian Bush, and Simon~C. Benjamin.
\newblock Quest and high performance simulation of quantum computers.
\newblock \emph{Scientific Reports}, 9:\penalty0 10736, 2019.
\newblock \doi{10.1038/s41598-019-47174-9}.

\bibitem[Bylaska et~al.(2021)Bylaska, Song, Bauman, Kowalski, Claudino, and
  Humble]{bylaska2021quantum}
Eric~J Bylaska, Duo Song, Nicholas~P Bauman, Karol Kowalski, Daniel Claudino,
  and Travis~S Humble.
\newblock Quantum solvers for plane-wave hamiltonians: Abridging virtual spaces
  through the optimization of pairwise correlations.
\newblock \emph{Frontiers In Chemistry}, 9:\penalty0 26, 2021.
\newblock \doi{doi.org/10.3389/fchem.2021.603019}.

\bibitem[Yalouz et~al.(2021)Yalouz, Senjean, G\"unther, Buda, O'Brien, and
  Visscher]{YalSenGun21}
Saad Yalouz, Bruno Senjean, Jakob G\"unther, Francesco Buda, Thomas~E O'Brien,
  and Lucas Visscher.
\newblock A state-averaged orbital-optimized hybrid
  quantum{\textendash}classical algorithm for a democratic description of
  ground and excited states.
\newblock \emph{Quantum Science and Technology}, 6\penalty0 (2):\penalty0
  024004, jan 2021.
\newblock \doi{10.1088/2058-9565/abd334}.

\bibitem[Manrique et~al.(2021)Manrique, Khan, Yamamoto, Wichitwechkarn, and
  Ramo]{ManKhaYam21}
David~Zsolt Manrique, Irfan~T. Khan, Kentaro Yamamoto, Vijja Wichitwechkarn,
  and David~Muñoz Ramo.
\newblock Momentum-space unitary coupled cluster and translational quantum
  subspace expansion for periodic systems on quantum computers.
\newblock \emph{arXiv:quant-ph}, 2008.08694, 2021.
\newblock \doi{10.48550/arXiv.2008.08694}.

\bibitem[Xia and Kais(2020)]{XiaKai20}
Rongxin Xia and Sabre Kais.
\newblock Qubit coupled cluster singles and doubles variational quantum
  eigensolver ansatz for electronic structure calculations.
\newblock \emph{Quantum Science and Technology}, 6\penalty0 (1):\penalty0
  015001, 2020.
\newblock \doi{10.1088/2058-9565/abbc74}.

\bibitem[Li et~al.(2021)Li, Huang, Cao, Huang, Shuai, et~al.]{LiLv2022}
Weitang Li, Zigeng Huang, Changsu Cao, Yifei Huang, Zhigang Shuai, et~al.
\newblock Toward practical quantum embedding simulation of realistic chemical
  systems on near-term quantum computers.
\newblock \emph{Chemical Science}, 13:\penalty0 8953--8962, 2021.
\newblock \doi{10.1039/D2SC01492K}.

\bibitem[Liu et~al.(2020)Liu, Wan, Li, and Yang]{LiuWanLi20}
Jie Liu, Lingyun Wan, Zhenyu Li, and Jinlong Yang.
\newblock Simulating periodic systems on a quantum computer using molecular
  orbitals.
\newblock \emph{J. Chem. Theory Comput.}, 16:\penalty0 6904--6914, 2020.
\newblock \doi{10.1021/acs.jctc.0c00881}.

\bibitem[Fan et~al.(2021{\natexlab{a}})Fan, Liu, Li, and Yang]{FanLiuLi21}
Yi~Fan, Jie Liu, Zhenyu Li, and Jinlong Yang.
\newblock Equation-of-motion theory to calculate accurate band structures with
  a quantum computer.
\newblock \emph{The Journal of Physical Chemistry Letters}, 12\penalty0
  (36):\penalty0 8833--8840, 2021{\natexlab{a}}.
\newblock \doi{doi.org/10.1021/acs.jpclett.1c02153}.

\bibitem[Kottmann et~al.(2021)Kottmann, Schleich, Tamayo-Mendoza, and
  Aspuru-Guzik]{KotSchTam21}
Jakob~S. Kottmann, Philipp Schleich, Teresa Tamayo-Mendoza, and Al\'an
  Aspuru-Guzik.
\newblock Reducing qubit requirements while maintaining numerical precision for
  the variational quantum eigensolver: A basis-set-free approach.
\newblock \emph{The Journal of Physical Chemistry Letters}, 12\penalty0
  (1):\penalty0 663--673, 2021.
\newblock \doi{doi.org/10.1021/acs.jpclett.0c03410}.

\bibitem[Cao et~al.(2022{\natexlab{a}})Cao, Hu, Zhang, Xu, Chen, Yu, Li, Hu,
  Lv, and Yung]{CaoHuZha21}
Changsu Cao, Jiaqi Hu, Wengang Zhang, Xusheng Xu, Dechin Chen, Fan Yu, Jun Li,
  Han-Shi Hu, Dingshun Lv, and Man-Hong Yung.
\newblock Progress toward larger molecular simulation on a quantum computer:
  Simulating a system with up to 28 qubits accelerated by point-group symmetry.
\newblock \emph{Physical Review A}, 105:\penalty0 062452, Jun
  2022{\natexlab{a}}.
\newblock \doi{10.1103/PhysRevA.105.062452}.

\bibitem[Ryabinkin et~al.(2021)Ryabinkin, Izmaylov, and Genin]{RyaIZmGen21}
Ilya~G Ryabinkin, Artur~F Izmaylov, and Scott~N Genin.
\newblock A posteriori corrections to the iterative qubit coupled cluster
  method to minimize the use of quantum resources in large-scale calculations.
\newblock \emph{Quantum Science and Technology}, 6\penalty0 (2):\penalty0
  024012, mar 2021.
\newblock \doi{10.1088/2058-9565/abda8e}.

\bibitem[Cao et~al.(2022{\natexlab{b}})Cao, Hu, Zhang, Xu, Chen, Yu, Li, Hu,
  Lv, and Yung]{CaoYung2022}
Changsu Cao, Jiaqi Hu, Wengang Zhang, Xusheng Xu, Dechin Chen, Fan Yu, Jun Li,
  Han-Shi Hu, Dingshun Lv, and Man-Hong Yung.
\newblock Progress toward larger molecular simulation on a quantum computer:
  Simulating a system with up to 28 qubits accelerated by point-group symmetry.
\newblock \emph{Phys. Rev. A}, 105:\penalty0 062452, Jun 2022{\natexlab{b}}.
\newblock \doi{10.1103/PhysRevA.105.062452}.

\bibitem[Hastings(2007)]{Hastings2007}
M~B Hastings.
\newblock An area law for one-dimensional quantum systems.
\newblock \emph{Journal of Statistical Mechanics: Theory and Experiment},
  2007\penalty0 (08):\penalty0 P08024--P08024, aug 2007.
\newblock \doi{10.1088/1742-5468/2007/08/p08024}.

\bibitem[McCaskey et~al.(2018)McCaskey, Dumitrescu, Chen, Lyakh, and
  Humble]{McCaskeyHumble2018}
Alexander McCaskey, Eugene Dumitrescu, Mengsu Chen, Dmitry Lyakh, and Travis
  Humble.
\newblock Validating quantum-classical programming models with tensor network
  simulations.
\newblock \emph{PLOS ONE}, 13\penalty0 (12):\penalty0 1--19, 12 2018.
\newblock \doi{10.1371/journal.pone.0206704}.

\bibitem[Zhou et~al.(2020)Zhou, Stoudenmire, and Waintal]{ZhouXavier2020}
Yiqing Zhou, E.~Miles Stoudenmire, and Xavier Waintal.
\newblock What limits the simulation of quantum computers?
\newblock \emph{Phys. Rev. X}, 10:\penalty0 041038, Nov 2020.
\newblock \doi{10.1103/PhysRevX.10.041038}.

\bibitem[Shang et~al.(2022)Shang, Shen, Fan, Xu, Guo, Liu, Zhou, Ma, Lin, Yang,
  Li, Wang, Zhang, and Li]{ShangLi2022}
Honghui Shang, Li~Shen, Yi~Fan, Zhiqian Xu, Chu Guo, Jie Liu, Wenhao Zhou, Huan
  Ma, Rongfen Lin, Yuling Yang, Fang Li, Zhuoya Wang, Yunquan Zhang, and Zhenyu
  Li.
\newblock {Large-Scale Simulation of Quantum Computational Chemistry on a New
  Sunway Supercomputer}.
\newblock \emph{arXiv:quant-ph}, 2207.03711, 2022.
\newblock \doi{10.48550/arXiv.2207.03711}.

\bibitem[Aleksandrowicz et~al.(2019)Aleksandrowicz, Alexander, Barkoutsos,
  Bello, Ben-Haim, Bucher, Cabrera-Hernández, Carballo-Franquis, Chen, Chen,
  Chow, Córcoles-Gonzales, Cross, Cross, Cruz-Benito, Culver, González,
  Torre, Ding, Dumitrescu, Duran, Eendebak, Everitt, Sertage, Frisch, Fuhrer,
  Gambetta, Gago, Gomez-Mosquera, Greenberg, Hamamura, Havlicek, Hellmers,
  Łukasz Herok, Horii, Hu, Imamichi, Itoko, Javadi-Abhari, Kanazawa, Karazeev,
  Krsulich, Liu, Luh, Maeng, Marques, Martín-Fernández, McClure, McKay,
  Meesala, Mezzacapo, Moll, Rodríguez, Nannicini, Nation, Ollitrault,
  O'Riordan, Paik, Pérez, Phan, Pistoia, Prutyanov, Reuter, Rice, Davila,
  Rudy, Ryu, Sathaye, Schnabel, Schoute, Setia, Shi, Silva, Siraichi,
  Sivarajah, Smolin, Soeken, Takahashi, Tavernelli, Taylor, Taylour, Trabing,
  Treinish, Turner, Vogt-Lee, Vuillot, Wildstrom, Wilson, Winston, Wood, Wood,
  Wörner, Akhalwaya, and Zoufal]{qiskit}
Gadi Aleksandrowicz, Thomas Alexander, Panagiotis Barkoutsos, Luciano Bello,
  Yael Ben-Haim, David Bucher, Francisco~Jose Cabrera-Hernández, Jorge
  Carballo-Franquis, Adrian Chen, Chun-Fu Chen, Jerry~M. Chow, Antonio~D.
  Córcoles-Gonzales, Abigail~J. Cross, Andrew Cross, Juan Cruz-Benito, Chris
  Culver, Salvador De La~Puente González, Enrique De~La Torre, Delton Ding,
  Eugene Dumitrescu, Ivan Duran, Pieter Eendebak, Mark Everitt, Ismael~Faro
  Sertage, Albert Frisch, Andreas Fuhrer, Jay Gambetta, Borja~Godoy Gago, Juan
  Gomez-Mosquera, Donny Greenberg, Ikko Hamamura, Vojtech Havlicek, Joe
  Hellmers, Łukasz Herok, Hiroshi Horii, Shaohan Hu, Takashi Imamichi,
  Toshinari Itoko, Ali Javadi-Abhari, Naoki Kanazawa, Anton Karazeev, Kevin
  Krsulich, Peng Liu, Yang Luh, Yunho Maeng, Manoel Marques, Francisco~Jose
  Martín-Fernández, Douglas~T. McClure, David McKay, Srujan Meesala, Antonio
  Mezzacapo, Nikolaj Moll, Diego~Moreda Rodríguez, Giacomo Nannicini, Paul
  Nation, Pauline Ollitrault, Lee~James O'Riordan, Hanhee Paik, Jesús Pérez,
  Anna Phan, Marco Pistoia, Viktor Prutyanov, Max Reuter, Julia Rice,
  Abdón~Rodríguez Davila, Raymond Harry~Putra Rudy, Mingi Ryu, Ninad Sathaye,
  Chris Schnabel, Eddie Schoute, Kanav Setia, Yunong Shi, Adenilton Silva,
  Yukio Siraichi, Seyon Sivarajah, John~A. Smolin, Mathias Soeken, Hitomi
  Takahashi, Ivano Tavernelli, Charles Taylor, Pete Taylour, Kenso Trabing,
  Matthew Treinish, Wes Turner, Desiree Vogt-Lee, Christophe Vuillot,
  Jonathan~A. Wildstrom, Jessica Wilson, Erick Winston, Christopher Wood,
  Stephen Wood, Stefan Wörner, Ismail~Yunus Akhalwaya, and Christa Zoufal.
\newblock {Qiskit: An Open-source Framework for Quantum Computing}, January
  2019.

\bibitem[Gray(2018)]{quimb}
Johnnie Gray.
\newblock quimb: A python package for quantum information and many-body
  calculations.
\newblock \emph{Journal of Open Source Software}, 3\penalty0 (29):\penalty0
  819, 2018.
\newblock \doi{10.21105/joss.00819}.

\bibitem[Mitarai et~al.(2018)Mitarai, Negoro, Kitagawa, and
  Fujii]{MitaraiFujii2018}
K.~Mitarai, M.~Negoro, M.~Kitagawa, and K.~Fujii.
\newblock Quantum circuit learning.
\newblock \emph{Phys. Rev. A}, 98:\penalty0 032309, Sep 2018.
\newblock \doi{10.1103/PhysRevA.98.032309}.

\bibitem[Cerezo et~al.(2021)Cerezo, Arrasmith, Babbush, Benjamin, Endo, Fujii,
  McClean, Mitarai, Yuan, Cincio, and Coles]{Cerezo2021}
M.~Cerezo, Andrew Arrasmith, Ryan Babbush, Simon~C. Benjamin, Suguru Endo,
  Keisuke Fujii, Jarrod~R. McClean, Kosuke Mitarai, Xiao Yuan, Lukasz Cincio,
  and Patrick~J. Coles.
\newblock {Variational quantum algorithms}.
\newblock \emph{Nature Reviews Physics}, 3\penalty0 (9):\penalty0 625--644,
  2021.
\newblock \doi{10.1038/s42254-021-00348-9}.

\bibitem[Sch\"on et~al.(2005)Sch\"on, Solano, Verstraete, Cirac, and
  Wolf]{SchonWolf2005}
C.~Sch\"on, E.~Solano, F.~Verstraete, J.~I. Cirac, and M.~M. Wolf.
\newblock Sequential generation of entangled multiqubit states.
\newblock \emph{Phys. Rev. Lett.}, 95:\penalty0 110503, Sep 2005.
\newblock \doi{10.1103/PhysRevLett.95.110503}.

\bibitem[Wei et~al.(2022)Wei, Malz, and Cirac]{WeiCirac2022}
Zhi-Yuan Wei, Daniel Malz, and J.~Ignacio Cirac.
\newblock Sequential generation of projected entangled-pair states.
\newblock \emph{Phys. Rev. Lett.}, 128:\penalty0 010607, Jan 2022.
\newblock \doi{10.1103/PhysRevLett.128.010607}.

\bibitem[Peruzzo et~al.(2014)Peruzzo, McClean, Shadbolt, Yung, Zhou, Love,
  Aspuru-Guzik, and O’Brien]{peruzzo_variational_2014}
Alberto Peruzzo, Jarrod McClean, Peter Shadbolt, Man-Hong Yung, Xiao-Qi Zhou,
  Peter~J. Love, Alán Aspuru-Guzik, and Jeremy~L. O’Brien.
\newblock A variational eigenvalue solver on a photonic quantum processor.
\newblock \emph{Nature Communications}, 5\penalty0 (1):\penalty0 4213, 2014.
\newblock \doi{doi.org/10.1038/ncomms5213}.

\bibitem[O’Malley et~al.(2016)O’Malley, Babbush, Kivlichan, Romero,
  McClean, Barends, Kelly, Roushan, Tranter, Ding, Campbell, Chen, Chen,
  Chiaro, Dunsworth, Fowler, Jeffrey, Lucero, Megrant, Mutus, Neeley, Neill,
  Quintana, Sank, Vainsencher, Wenner, White, Coveney, Love, Neven,
  Aspuru-Guzik, and Martinis]{omalley_scalable_2016}
P.~J.~J. O’Malley, R.~Babbush, I.~D. Kivlichan, J.~Romero, J.~R. McClean,
  R.~Barends, J.~Kelly, P.~Roushan, A.~Tranter, N.~Ding, B.~Campbell, Y.~Chen,
  Z.~Chen, B.~Chiaro, A.~Dunsworth, A.~G. Fowler, E.~Jeffrey, E.~Lucero,
  A.~Megrant, J.~Y. Mutus, M.~Neeley, C.~Neill, C.~Quintana, D.~Sank,
  A.~Vainsencher, J.~Wenner, T.~C. White, P.~V. Coveney, P.~J. Love, H.~Neven,
  A.~Aspuru-Guzik, and J.~M. Martinis.
\newblock Scalable quantum simulation of molecular energies.
\newblock \emph{Physical Review X}, 6\penalty0 (3):\penalty0 031007, 2016.
\newblock \doi{doi.org/10.1103/PhysRevX.6.031007}.

\bibitem[Kandala et~al.(2017)Kandala, Mezzacapo, Temme, Takita, Brink, Chow,
  and Gambetta]{kandala_hardware-efficient_2017}
Abhinav Kandala, Antonio Mezzacapo, Kristan Temme, Maika Takita, Markus Brink,
  Jerry~M. Chow, and Jay~M. Gambetta.
\newblock Hardware-efficient variational quantum eigensolver for small
  molecules and quantum magnets.
\newblock \emph{Nature}, 549\penalty0 (7671):\penalty0 242--246, 2017.
\newblock \doi{doi.org/10.1038/nature23879}.

\bibitem[Colless et~al.(2018)Colless, Ramasesh, Dahlen, Blok, Kimchi-Schwartz,
  McClean, Carter, de~Jong, and Siddiqi]{colless_computation_2018}
J.~I. Colless, V.~V. Ramasesh, D.~Dahlen, M.~S. Blok, M.~E. Kimchi-Schwartz,
  J.~R. McClean, J.~Carter, W.~A. de~Jong, and I.~Siddiqi.
\newblock Computation of molecular spectra on a quantum processor with an
  error-resilient algorithm.
\newblock \emph{Physical Review X}, 8\penalty0 (1):\penalty0 011021, 2018.
\newblock \doi{doi.org/10.1103/PhysRevX.8.011021}.

\bibitem[Sagastizabal et~al.(2019)Sagastizabal, Bonet-Monroig, Singh, Rol,
  Bultink, Fu, Price, Ostroukh, Muthusubramanian, Bruno, Beekman, Haider,
  O'Brien, and DiCarlo]{expri-symmetry-1}
R.~Sagastizabal, X.~Bonet-Monroig, M.~Singh, M.~A. Rol, C.~C. Bultink, X.~Fu,
  C.~H. Price, V.~P. Ostroukh, N.~Muthusubramanian, A.~Bruno, M.~Beekman,
  N.~Haider, T.~E. O'Brien, and L.~DiCarlo.
\newblock Experimental error mitigation via symmetry verification in a
  variational quantum eigensolver.
\newblock \emph{Phys. Rev. A}, 100:\penalty0 010302, 2019.
\newblock \doi{doi.org/10.1103/PhysRevA.100.010302}.

\bibitem[Shen et~al.(2017)Shen, Zhang, Zhang, Zhang, Yung, and
  Kim]{expri-vqe-1}
Yangchao Shen, Xiang Zhang, Shuaining Zhang, Jing-Ning Zhang, Man-Hong Yung,
  and Kihwan Kim.
\newblock Quantum implementation of the unitary coupled cluster for simulating
  molecular electronic structure.
\newblock \emph{Phys. Rev. A}, 95:\penalty0 020501, 2017.
\newblock \doi{doi.org/10.1103/PhysRevA.95.020501}.

\bibitem[Hempel et~al.(2018)Hempel, Maier, Romero, McClean, Monz, Shen,
  Jurcevic, Lanyon, Love, Babbush, Aspuru-Guzik, Blatt, and
  Roos]{hempel_quantum_2018}
Cornelius Hempel, Christine Maier, Jonathan Romero, Jarrod McClean, Thomas
  Monz, Heng Shen, Petar Jurcevic, Ben~P. Lanyon, Peter Love, Ryan Babbush,
  Alan Aspuru-Guzik, Rainer Blatt, and Christian~F. Roos.
\newblock Quantum chemistry calculations on a trapped-ion quantum simulator.
\newblock \emph{Physical Review X}, 8\penalty0 (3):\penalty0 031022, 2018.
\newblock \doi{doi.org/10.1103/PhysRevX.8.031022}.

\bibitem[Nam et~al.(2019)Nam, Chen, Pisenti, Wright, Delaney, Maslov, Brown,
  Allen, Amini, Apisdorf, Beck, Blinov, Chaplin, Chmielewski,
  et~al.]{nam_ground-state_2019}
Yunseong Nam, Jwo-Sy Chen, Neal~C. Pisenti, Kenneth Wright, Conor Delaney,
  Dmitri Maslov, Kenneth~R. Brown, Stewart Allen, Jason~M. Amini, Joel
  Apisdorf, Kristin~M. Beck, Aleksey Blinov, Vandiver Chaplin, Mika
  Chmielewski, et~al.
\newblock Ground-state energy estimation of the water molecule on a trapped ion
  quantum computer.
\newblock \emph{npj Quantum Information}, 6\penalty0 (1):\penalty0 33, 2019.
\newblock \doi{doi.org/10.1038/s41534-020-0259-3}.

\bibitem[Romero et~al.(2018)Romero, Babbush, McClean, Hempel, Love, and
  Aspuru-Guzik]{UCCansatz}
Jonathan Romero, Ryan Babbush, Jarrod~R McClean, Cornelius Hempel, Peter~J
  Love, and Al{\'{a}}n Aspuru-Guzik.
\newblock Strategies for quantum computing molecular energies using the unitary
  coupled cluster ansatz.
\newblock \emph{Quantum Science and Technology}, 4\penalty0 (1):\penalty0
  014008, oct 2018.
\newblock \doi{10.1088/2058-9565/aad3e4}.

\bibitem[Babbush et~al.(2015)Babbush, McClean, Wecker,
  et~al.]{Babbush_Trotter_2015}
Ryan Babbush, Jarrod McClean, Dave Wecker, et~al.
\newblock Chemical basis of trotter-suzuki errors in quantum chemistry
  simulation.
\newblock \emph{Phys. Rev. A}, 91:\penalty0 022311, 2015.
\newblock \doi{10.1103/PhysRevA.91.022311}.

\bibitem[Grimsley et~al.(2020)Grimsley, Claudino, Economou,
  et~al.]{Grimsley_Trotter_2020}
Harper~R. Grimsley, Daniel Claudino, Sophia~E. Economou, et~al.
\newblock Is the trotterized uccsd ansatz chemically well-defined?
\newblock \emph{J.~Chem.\ Theory Comput.}, 16:\penalty0 1--6, 2020.
\newblock \doi{10.1021/acs.jctc.9b01083}.

\bibitem[Guo et~al.(2019{\natexlab{a}})Guo, Liu, Xiong, Xue, Fu, Huang, Qiang,
  Xu, Liu, Zheng, Huang, Deng, Poletti, Bao, and Wu]{GuoChu2019}
Chu Guo, Yong Liu, Min Xiong, Shichuan Xue, Xiang Fu, Anqi Huang, Xiaogang
  Qiang, Ping Xu, Junhua Liu, Shenggen Zheng, He-Liang Huang, Mingtang Deng,
  Dario Poletti, Wan-Su Bao, and Junjie Wu.
\newblock General-purpose quantum circuit simulator with projected
  entangled-pair states and the quantum supremacy frontier.
\newblock \emph{Phys. Rev. Lett.}, 123:\penalty0 190501, Nov
  2019{\natexlab{a}}.
\newblock \doi{10.1103/PhysRevLett.123.190501}.

\bibitem[Vidal(2004)]{Vidal2004}
Guifr\'e Vidal.
\newblock Efficient simulation of one-dimensional quantum many-body systems.
\newblock \emph{Phys. Rev. Lett.}, 93:\penalty0 040502, Jul 2004.
\newblock \doi{10.1103/PhysRevLett.93.040502}.

\bibitem[Hastings(2009)]{Hastings2009}
Matthew~B Hastings.
\newblock Light-cone matrix product.
\newblock \emph{Journal of mathematical physics}, 50\penalty0 (9):\penalty0
  095207, 2009.
\newblock \doi{doi.org/10.1063/1.3149556}.

\bibitem[Griewank(1992)]{autodiff}
Andreas Griewank.
\newblock Achieving logarithmic growth of temporal and spatial complexity in
  reverse automatic differentiation.
\newblock \emph{Optimization Methods and Software}, 1\penalty0 (1):\penalty0
  35--54, 1992.
\newblock \doi{10.1080/10556789208805505}.

\bibitem[Guo and Poletti(2021)]{GuoPoletti2021}
Chu Guo and Dario Poletti.
\newblock Scheme for automatic differentiation of complex loss functions with
  applications in quantum physics.
\newblock \emph{Phys. Rev. E}, 103:\penalty0 013309, Jan 2021.
\newblock \doi{10.1103/PhysRevE.103.013309}.

\bibitem[Luo et~al.(2020)Luo, Liu, Zhang, and Wang]{Yao}
Xiu-Zhe Luo, Jin-Guo Liu, Pan Zhang, and Lei Wang.
\newblock Yao.jl: {E}xtensible, {E}fficient {F}ramework for {Q}uantum
  {A}lgorithm {D}esign.
\newblock \emph{{Quantum}}, 4:\penalty0 341, October 2020.
\newblock ISSN 2521-327X.
\newblock \doi{10.22331/q-2020-10-11-341}.

\bibitem[Liao et~al.(2019)Liao, Liu, Wang, and Xiang]{LiaoXiang2019}
Hai-Jun Liao, Jin-Guo Liu, Lei Wang, and Tao Xiang.
\newblock Differentiable programming tensor networks.
\newblock \emph{Phys. Rev. X}, 9:\penalty0 031041, Sep 2019.
\newblock \doi{10.1103/PhysRevX.9.031041}.

\bibitem[McClean et~al.(2018)McClean, Boixo, Smelyanskiy, Babbush, and
  Neven]{McCleanNeven2018}
Jarrod~R McClean, Sergio Boixo, Vadim~N Smelyanskiy, Ryan Babbush, and Hartmut
  Neven.
\newblock Barren plateaus in quantum neural network training landscapes.
\newblock \emph{Nature communications}, 9\penalty0 (1):\penalty0 4812, 2018.
\newblock \doi{10.1038/s41467-018-07090-4}.

\bibitem[Powell(2009)]{Powell_BOBYQA_2009}
M.~Powell.
\newblock The bobyqa algorithm for bound constrained optimization without
  derivatives.
\newblock \emph{Technical Report, Department of Applied Mathematics and
  Theoretical Physics}, 01 2009.

\bibitem[Jor(2006)]{Jorge_OPT_1999}
\emph{Quasi-Newton Methods}, pages 135--163.
\newblock Springer New York, New York, NY, 2006.
\newblock ISBN 978-0-387-40065-5.
\newblock \doi{10.1007/978-0-387-40065-5_6}.

\bibitem[Sapova and Fedorov(2022)]{Sapova_CO2_2022}
Mariia~D. Sapova and Aleksey~K. Fedorov.
\newblock Variational quantum eigensolver techniques for simulating carbon
  monoxide oxidation.
\newblock \emph{Communications Physics}, 5\penalty0 (1):\penalty0 199, Aug
  2022.
\newblock ISSN 2399-3650.
\newblock \doi{10.1038/s42005-022-00982-4}.

\bibitem[Fan et~al.(2021{\natexlab{b}})Fan, Cao, Xu, Li, Lv, and Yung]{ES_2021}
Yi~Fan, Changsu Cao, Xusheng Xu, Zhenyu Li, Dingshun Lv, and Man-Hong Yung.
\newblock Circuit-depth reduction of unitary-coupled-cluster ansatz by energy
  sorting.
\newblock \emph{arXiv:quant-ph}, 2106.15210, 2021{\natexlab{b}}.
\newblock \doi{10.48550/arXiv.2106.15210}.

\bibitem[Lee et~al.(2019)Lee, Huggins, Head-Gordon, and Whaley]{uccgsd}
Joonho Lee, William~J. Huggins, Martin Head-Gordon, and K.~Birgitta Whaley.
\newblock Generalized unitary coupled cluster wave functions for quantum
  computation.
\newblock \emph{Journal of Chemical Theory and Computation}, 15\penalty0
  (1):\penalty0 311--324, 2019.
\newblock \doi{10.1021/acs.jctc.8b01004}.

\bibitem[Grimsley et~al.(2019)Grimsley, Economou, Barnes, and
  Mayhall]{ADAPT-VQE}
Harper~R. Grimsley, Sophia~E. Economou, Edwin Barnes, and Nicholas~J. Mayhall.
\newblock An adaptive variational algorithm for exact molecular simulations on
  a quantum computer.
\newblock \emph{Nature Communications}, 10\penalty0 (1), jul 2019.
\newblock \doi{10.1038/s41467-019-10988-2}.

\bibitem[Sun et~al.(2018)Sun, Berkelbach, Blunt, Booth, Guo, Li, Liu, McClain,
  Sayfutyarova, Sharma, Wouters, and Chan]{pyscf}
Qiming Sun, Timothy~C. Berkelbach, Nick~S. Blunt, George~H. Booth, Sheng Guo,
  Zhendong Li, Junzi Liu, James~D. McClain, Elvira~R. Sayfutyarova, Sandeep
  Sharma, Sebastian Wouters, and Garnet Kin-Lic Chan.
\newblock Pyscf: the python-based simulations of chemistry framework.
\newblock \emph{WIREs Computational Molecular Science}, 8\penalty0
  (1):\penalty0 e1340, 2018.
\newblock \doi{10.1002/wcms.1340}.

\bibitem[Guo et~al.(2019{\natexlab{b}})Guo, Liu, Xiong, Xue, Fu, Huang, Qiang,
  Xu, Liu, Zheng, Huang, Deng, Poletti, Bao, and Wu]{GuoWu2019}
Chu Guo, Yong Liu, Min Xiong, Shichuan Xue, Xiang Fu, Anqi Huang, Xiaogang
  Qiang, Ping Xu, Junhua Liu, Shenggen Zheng, He-Liang Huang, Mingtang Deng,
  Dario Poletti, Wan-Su Bao, and Junjie Wu.
\newblock General-purpose quantum circuit simulator with projected
  entangled-pair states and the quantum supremacy frontier.
\newblock \emph{Phys. Rev. Lett.}, 123:\penalty0 190501, Nov
  2019{\natexlab{b}}.
\newblock \doi{10.1103/PhysRevLett.123.190501}.

\bibitem[Hikihara et~al.(2023)Hikihara, Ueda, Okunishi, Harada, and
  Nishino]{HikiharaNishino2023}
Toshiya Hikihara, Hiroshi Ueda, Kouichi Okunishi, Kenji Harada, and Tomotoshi
  Nishino.
\newblock Automatic structural optimization of tree tensor networks.
\newblock \emph{Phys. Rev. Res.}, 5:\penalty0 013031, Jan 2023.
\newblock \doi{10.1103/PhysRevResearch.5.013031}.

\bibitem[Bezanson et~al.(2017)Bezanson, Edelman, Karpinski,
  et~al.]{Bezanson_JULIA_2017}
Jeff Bezanson, Alan Edelman, Stefan Karpinski, et~al.
\newblock Julia: A fresh approach to numerical computing.
\newblock \emph{SIAM Review}, 59\penalty0 (1):\penalty0 65--98, 2017.
\newblock \doi{10.1137/141000671}.

\bibitem[Byrne et~al.(2021)Byrne, Wilcox, and Churavy]{byrne2021mpi}
Simon Byrne, Lucas~C Wilcox, and Valentin Churavy.
\newblock Mpi. jl: Julia bindings for the message passing interface.
\newblock In \emph{Proceedings of the JuliaCon Conferences}, volume~1, page~68,
  2021.

\bibitem[Guo(2022)]{diffMPS}
Chu Guo.
\newblock \text{MPSSimulator}.
\newblock \emph{GitHub Repository}, 2022.

\bibitem[Johnson(2007)]{Steven_NLopt_2007}
Steven~G. Johnson.
\newblock The {NLopt} nonlinear-optimization package.
\newblock \emph{GitHub Repository}, 2007.

\end{thebibliography}

\end{document}